\documentstyle[11pt]{article}
\def\IR{\relax{\rm I\kern-.18em R}}

\def\g{\gamma}
\newcommand{\be}{\begin{equation}}
\newcommand{\ee}{\end{equation}}
\newcommand{\bea}{\begin{eqnarray}}
\newcommand{\eea}{\end{eqnarray}}
\let\bm=\bibitem

\def\r{\rho}

\def\o{\omega}

\def\4{{\sst{(4)}}}

\begin{document}
\pagestyle{empty}

\vspace{3.0cm}

\

\

\

\centerline{\Large \bf Beta, Dipole and Noncommutative Deformations of M-theory}
\vspace{0.3cm}
\centerline{\Large \bf  Backgrounds with One or More Parameters}
\vspace{1.5cm}
\centerline{Aybike \c{C}atal-\"{O}zer$^1$ and Nihat Sadik Deger$^{2,3}$}

\

$^1$ Dept. of Mathematics, Istanbul Technical University, Maslak, 34469, Istanbul-Turkey \\
$^2$ Dept. of Mathematics, Bogazici University, Bebek, 34342, Istanbul-Turkey \\
$^3$ Feza Gursey Institute, Cengelkoy, 34680, Istanbul-Turkey
\

{\bf E-mails:} ozerayb@itu.edu.tr, \, sadik.deger@boun.edu.tr

\vspace{1.5cm}

\centerline{\bf ABSTRACT}
\vspace{0.5cm}
We construct new M-theory solutions starting from those
that contain  5 $U(1)$ isometries. We do this by reducing along
one of the 5-torus directions, then T-dualizing via the action of
an $O(4,4)$ matrix and lifting back to 11-dimensions. The
particular T-duality transformation is a sequence of $O(2,2)$
transformations embedded in $O(4,4)$, where the action of each
$O(2,2)$ gives a Lunin-Maldacena deformation in 10-dimensions. We
find general formulas for the metric and 4-form field of single
and multiparameter deformed solutions, when the 4-form of the
initial 11-dimensional background has at most one leg along the
$5$-torus. All the deformation terms in the new solutions are
given in terms of subdeterminants of a $5 \times 5$ matrix,  which
represents the metric on the $5$-torus.  We apply these results to
several M-theory backgrounds of the type $AdS_r \times X^{11-r}$.
By appropriate choices of the T-duality and reduction directions
we obtain analogues of beta, dipole and noncommutative
deformations. We also provide formulas for backgrounds with only 3
or 4 $U(1)$ isometries and study a case, for which our assumption
for the 4-form field is violated.

\vspace{2cm}

{\bf Keywords:} AdS-CFT Correspondence, M-theory, String Duality

{\bf PACS:} 11.25.Yb, 11.25.Tq

\newpage

\pagestyle{plain}

\tableofcontents

\section{Introduction}

Construction of new M-theory solutions has been an important area
of research for a long time. The goal  of this paper is to
generate new M-theory solutions by deforming those that involve
five $U(1)$ isometries. Our method is a generalization of the
Lunin-Maldacena procedure \cite{mal2}, which gives the string
theory duals of $\beta$ deformations of certain field theories.

$\beta$ deformations can be applied to $U(N)$ field theories with
a $U(1) \times U(1)$ global symmetry. An example is the
Leigh-Strassler deformation \cite{leigh} of the $N=4$ Super
Yang-Mills theory, which breaks the supersymmetry to $N=1$.  The
global $U(1) \times U(1)$ symmetry of the field theory corresponds
to a 2-torus in the dual gravity picture. A string theory
background with a 2-torus in its geometry possesses an $O(2,2)$
T-duality symmetry. Strictly speaking, the T-duality symmetry of
the string theory background is $O(2,2,Z)$, which
transforms the conformal field theory on the string world-sheet to
an equivalent one. On the other hand, the corresponding
supergravity theory has an $O(2,2,R)$ solution generating
symmetry, which transforms the conformal field theory on the
world-sheet to an exactly marginal deformation of it (see
\cite{T-duality1} for a review). There is the well-known
isomorphism $SO(2,2,R) \simeq SL(2,R)_{\tau}
\times SL(2,R)_{\rho}$, where the first factor acts on
the complex structure modulus and the second factor acts on the
K\"{a}hler modulus of the 2-torus. Lunin and Maldacena (LM)
\cite{mal2} used the latter to generate new Type IIB solutions and
showed that these solutions correspond to the $\beta$ deformation
of the field theory duals of the initial gravity solution, when
the 2-torus lies in the geometry in a certain way. Later in
\cite{aybike} the $O(2,2)$ matrix whose action generates the
deformed solutions was identified by considering the way
$SL(2,R)_{\rho}$ sits in $O(2,2,R)$. The method
of LM works for any 2-torus that lies in the background geometry
and by different choices one can obtain the duals of
noncommutative and dipole deformations, as well. Therefore, this
procedure provides a unified framework for studying different
types of deformations. In this paper, we refer to any of these as
a LM deformation.

This idea was generalized to construct new M-theory solutions by
deforming M-theory backgrounds which involve a $U(1)^3$ isometry
in their geometry \cite{mal2}. To do this, one can reduce along
one of the coordinates of the 3-torus to obtain a Type IIA
solution, use the $SL(2,R)_{\rho}$ symmetry associated
with the remaining two legs to generate a new solution in ten
dimensions and  lift back to eleven dimensions. The first eleven
dimensional example had already appeared in \cite{mal2}, where a
$\beta$ deformation of the $AdS_4 \times S^7$ solution was
obtained by using a 3-torus lying completely in $S^7$.  Later,
$\beta$ deformations of backgrounds of the type $AdS_4 \times
M_7$, where $M_7$ is a 7-dimensional Sasaki-Einstein manifold
\cite{sasaki1}-\cite{sasaki7} were performed in \cite{ahn} and
\cite{jerome}. Deformations of the membrane \cite{2brane} and
five-brane \cite{5brane} solutions of the D=11 supergravity
\cite{11} and their near horizon geometries were obtained in
\cite{berman}, where in addition to the $\beta$ deformations,
dipole and non-commutative deformations were considered, as well.
Recently, $\beta$ deformations of $AdS_4 \times S^7/Z_k$
were studied in \cite{imeroni}. All these are single parameter
deformations. All the backgrounds that have been considered so far
have a common feature: they all involve more than three (in fact,
five) $U(1)$ isometries, allowing a generalization of the LM
procedure. In this paper, we study  this generalization, and hence
construct new deformations involving one and more parameters.
Multiparameter deformations of ten dimensional string backgrounds
were first studied in \cite{frolov}.

Our method will be as follows. For a general eleven dimensional
background with $n \geq 3$ $U(1)$ isometries, we start by reducing
along one of the legs of the $n$-torus associated with these,
thereby obtaining a IIA solution with an $O(n-1,n-1)$ solution
generating symmetry. We deform this solution through the action of
an $O(n-1,n-1)$ matrix and lift back to eleven dimensions. Here, a
crucial (but not restrictive for the examples that are of interest
to us) assumption is that this $n$-torus should be decoupled from
the rest of the geometry. The particular T-duality transformation
we use is a sequence of $O(2,2)$ transformations embedded in
$O(n-1,n-1)$, where the action of each $O(2,2)$ gives a LM
deformation. Our procedure allows up to $n! / 6 (n-3)! $
parameters, which corresponds to the number of ways one can choose
3-dimensional subtori from the $n$-torus. When we present our
method in the next section, we choose $n=5$, and show that the
deformed solutions are of a universal form and all the terms
depending on the deformation parameter can be written in terms of
subdeterminants of a $5 \times 5 $ matrix, representing the metric
on the 5-torus. The choice $n=5$ is preferred basically for two
reasons. Firstly, all the examples that are of interest to us have
5 $U(1)$ isometries. Secondly, the general formulas obtained in
this case also includes the $n=3$ and $n=4$ cases, after an
additional assumption.

A one-parameter deformation obtained with our method (which is the
only possibility when $n=3$) corresponds to a choice of a
3-dimensional subtorus of the 5-torus in the geometry and gives an
ordinary LM deformation in 11 dimensions. However, even in this
simplest case, our method has the virtue of providing  general
formulas, which make the calculations much easier. The real
novelty arises when $n > 3$, which allows the introduction of more
than one deformation parameters. We illustrate our method through
several examples, all of which are of the form $AdS_r \times
X^{11-r}$ motivated, of course by the AdS/CFT correspondence
\cite{mal1, ads2, ads3}. Choosing the 3-torus to lie completely in
the compact part, completely in the noncompact part or partly in
the compact and partly in the noncompact part of the geometry
correspond to the analogues of the duals of $\beta$,
noncommutative and dipole deformations in string theory,
respectively. We also introduce ``mixed deformations", which are
multiparameter deformations involving several 3-tori, where each
3-torus gives rise to a different type of deformation.

The organization of our paper is as follows. In the next section
we describe our method for the simpler case of one parameter
deformations. We start by imposing rather mild conditions on the
4-form and the metric of a given 11 dimensional background and
derive general formulas for the deformed 4-form
(\ref{deformedF11asil}) and the metric (\ref{esassonuc}). Then we
apply these to the backgrounds $AdS_4 \times$
(Sasaki-Einstein)$_7$ (with base $CP^2$) \cite{sasaki2}
and $AdS_7 \times S^4$  in subsections 2.1 and 2.2. In subsection
2.3 we explain, through the $AdS_7 \times S^4$ example, how  our
method should be modified, when our assumption for the 4-form does
not hold. In section 3, we generalize our discussion to the
multiparameter case. After giving general formulas for the
deformed metric (\ref{3pesassonuc}) and the 4-form
(\ref{3pdeformedF11asil}), we obtain the 3-parameter $\beta$,
dipole and mixed deformations of the $AdS_4 \times$
(Sasaki-Einstein)$_7$ (with base $S^2 \times S^2$) solution
\cite{sasaki2}. We conclude with some comments and future
directions in section 4. Appendices contain proofs of two
equations which are used in deriving the general formulas and necessary subdeterminants for section 3.

\section{One Parameter Deformations}

In this section we  obtain the 1-parameter LM deformation of a
general eleven dimensional background with three or more
isometries.  Before we start, let us fix our notation. Throughout
the paper the hatted fields refer to eleven dimensional fields,
whereas fields without hats are in ten dimensions. We use tilde
for fields after deformation in ten or eleven dimensions. Our
index conventions are such that (unless otherwise indicated) $M,N$
run from 1 to 11, the indices $m,n,p,q,r$ count the five isometry
directions, running from 1 to 5, whereas $\mu, \nu $ count the
remaining coordinates from 6 to 11. We also have that $i,j,k,l \in
\{1,2,3,4\}$ and $a,b,c \in \{1,2,3\}$. For a general matrix $A$,
$A(i \mid j)$ denotes the matrix obtained from $A$ by deleting the
$i$th column and $j$th row. Similarly, $A(a,b \mid i,j)$ is the
matrix obtained from $A$ by deleting columns $a,b$ and rows $i,
j$. Note that when $A$ is symmetric  $det A(i \mid j) = det A(j
\mid i)$ and $det A(a,b \mid i,j)=  det A(i,j \mid a,b)$.

Before focusing on eleven dimensional backgrounds, let us review
the LM deformations in ten dimensions, where our approach will be
that of \cite{aybike}. LM deformations can be applied to
backgrounds with two $U(1)$ isometries. Let us label the
coordinates such that these isometries have Killing vectors
$\partial/\partial x^1$, $\partial / \partial x^2$. Suppose that
$x^1$ and/or $x^2$ couple to $d-2$ other coordinates, which we
label as $x^3, \cdots, x^d$. Then the deformed solution can be
expressed in a simple way using the so called background matrix
defined as: \be \label{background} E= g_{ij} + B_{ij} \, , \ \ \
i,j = 1, \cdots, d \ee where $g$ and $B$ are the matrices with
entries $ g_{ij},$ and $B_{ij}$. Here, $g_{ij}$ and $B_{ij}$ are
the components of the metric and the B-field of the background,
respectively. It was shown in \cite{aybike} that the deformed
solutions can be obtained via the action of the following $O(d,d)$
matrix: \be \label{matrix} T_d = \left(\begin{array}{ccc}
  a & & b \\
 c & & d \\
\end{array}\right) =\left(\begin{array}{ccc}
  1_d & & 0_d \\
 \Gamma_d & & 1_d \\
\end{array}\right),
\ee where $1_d$ and $0_d$ are the  $d \times d$ identity and null
matrices, respectively and $\Gamma_d$ is the $ d \times d$ matrix
of the form \be \label{matrixc} \Gamma_d =
\left(\begin{array}{ccccc}
  0 & - \gamma & 0 & \cdots & 0 \\
  \gamma & \ \ 0 & 0 & \cdots & 0 \\
  0 & \ 0 & 0 & \cdots & 0 \\
  \vdots & & & & \vdots \\
  0 & \ 0 & 0 & \cdots & 0
\end{array}\right).
\ee  $\gamma$ is a real constant. Here, the solution generating
symmetry is $O(2,2)$. However, the  associated T-duality matrix
$T^2$ has to be embedded in $O(d,d)$, due to the coupling of $x^1$
and/or $x^2$ with the coordinates $x^3, \cdots, x^d$
\cite{giveon}.

The transformation of the background matrix and the dilaton under
the action of the above $O(d,d)$ transformation is \cite{giveon}:
\bea \label{dilaton} e^{2\phi} &\longrightarrow&
e^{2\tilde{\phi}}=\frac{e^{2\phi}}{{\rm det}(\Gamma_d E + 1_d)} \, , \\
\label{onemli} E &\longrightarrow&  {\tilde E}= (a E + b)(c E +
d)^{-1}=E(\Gamma_d E + 1_d)^{-1} \, . \eea The other $g$ and $B$
components of the background do not get any additive $\gamma$
corrections.

The Ramond-Ramond fields transform in the spinorial representation
of $O(d,d)$. One can show that their transformation under the
particular $O(d,d)$ element (\ref{matrix}) can be found through
the action of the operator (see \cite{rr1, rr2, rr3, rr4} for
details)
\begin{equation}\label{RR7}
{\bf T} = \exp[\frac{1}{2} (\Gamma_d)_{mn} i_m i_n],
\end{equation}
where $i_m$ is contraction with respect to the isometry direction
$\partial/\partial x^m$, $i_m \equiv i_{\partial/\partial x^m}$.
The operator ${\bf T}$ acts on $ F$, which is defined as
\begin{equation}\label{RR2}
F \equiv e^{-B} \sum_{p=1}^9 F_p \, ,
\end{equation}
where $F_p$'s are the $p$-form field strengths with $p$ even in
type IIA and odd in type IIB theory. $F_p$ for $p > 5$ is defined
via its Hodge dual as $F_{10-p} = (-1)^{[\frac{p-1}{2}]} * F_p$,
where $[\frac{p-1}{2}]$ is the first integer greater than or equal
to $\frac{p-1}{2}$ \cite{ddII}. Using (\ref{RR7}) and
(\ref{matrixc}) we have \be \sum_{p=1}^9 \tilde{F}_p \wedge
e^{-\tilde{B}} = (1 - \gamma i_1 i_2) (\sum_{p=1}^9 F_p \wedge
e^{-B}) , \ee where $\tilde{F}_p$ and $\tilde{B}$ are fields after
the deformation. Then it follows that the transformation rule for
the 4-form and 2-form fields $F_4$ and $F_2$ in Type IIA theory
are \cite{imeroni, aybike}: \bea \label{f2f4} F_2
&\longrightarrow& \tilde{F}_2 = F_2 - \gamma i_{1} i_{2}(F_4 -
F_2 \wedge B)  \\
F_4 &\longrightarrow& \tilde{F}_4 = F_4 - F_2 \wedge B -\gamma
i_{1} i_{2}(F_6 - F_4 \wedge B + \frac{1}{2}F_2 \wedge B \wedge B)
+ \tilde{F}_2 \wedge \tilde{B}. \nonumber \eea

After this brief review of the method used in \cite{aybike} to
obtain the LM deformations, let us describe our method to deform a
given eleven dimensional background with three or more commuting
isometries. We start by singling out a 3-torus associated with
three of these isometries. Then we reduce to ten dimensions along
one of the directions of the 3-torus. We obtain a new solution in
ten dimensions by the action of the T-duality matrix
(\ref{matrix}), corresponding to the remaining two isometries.
Lifting back to eleven dimensions, we obtain the deformed M-theory
backgrounds. We make the following two assumptions on the given
11-dimensional solution:

\noindent $(i)$ Its metric contains $n \geq 3$ commuting
isometries, which decouple from other coordinates.

\noindent $(ii)$ Its 4-form field strength has at most one leg
along these $n$ directions.

We start by taking $n=5$ for convenience, as this will allow us to
discuss different types of deformations simultaneously. Moreover,
this is not a strong restriction, since many widely studied
M-theory backgrounds have this property, as we will see. However,
let us remark that we don't need all these 5 directions to
correspond to $U(1)$ isometries, only 3 will be enough to obtain a
single parameter deformation with our method. The remaining two
can be any directions that couple to the deformation 3-torus in
the metric. The cases in which there are no such couplings or the
background does not contain five $U(1)$ isometries will be
considered at the end of this section. The second assumption above
is required in order to give general formulas for the deformation
of the 4-form field, as we explain below. Our formulas will
be valid for \emph{any} given eleven dimensional background
meeting these two conditions. We will discuss how our method
works, when the second assumption is violated through an example
in subsection 2.3.

The standard ansatz for reducing a given $D=11$ background to 10
dimensions is \be \label{ansatz} ds_{11}^2 = \hat{g}_{MN} dx^M
dx^N = e^{-2/3 \phi} ds_{10 \ s}^2 + e^{4/3 \phi} (dz + A)^2 \ \ \
\ M,N = 1, \cdots,11 \ee Here $z$ is the coordinate along which we
are reducing, $A$ is a one-form and $\phi$ is the dilaton. The
subscript $s$ refers to the fact that the ten dimensional metric
is in the string frame.

We label the five $U(1)$ coordinates as $x^1, \cdots, x^5$. Our
first assumption implies that the initial eleven dimensional
metrics will be such that $\hat{g}_{m \mu} = 0$ for all
$m=1,\cdots,5 \ \ \ \mu = 6, \cdots, 11$, that is, these 5
isometry directions in the geometry is separated from the
transverse part to it. Then, choosing one of the isometry
coordinates as the coordinate $z$ along which we reduce, the
eleven dimensional metric using the reduction ansatz
(\ref{ansatz}) takes the form \bea \label{metric11a} ds^2_{11} &=&
\hat{g}_{\mu \nu} dx^{\mu} dx^{\nu} + \hat{g}_{mn} dx^m
dx^n, \nonumber \\
&=&  e^{-2/3\phi}(g_{\mu \nu} dx^{\mu} dx^{\nu} + g_{ij} dx^i
dx^j) + e^{4/3 \phi} (dz + A_i(x^{\mu}) dx^i)^2  \, ,  \ \ \  \eea
where $ m,n = 1,\cdots,5, \ \ \mu, \nu = 6, \cdots, 11, \ \ i,j =
1,\cdots,4$. Note that we have labeled $x^5 \equiv z$. One can
easily see that the fields with and without hats are related as
\bea \label{hatbar} g_{\mu \nu} &=& e^{2/3 \phi} \ \hat{g}_{\mu
\nu} = \sqrt{\hat{g}_{zz}} \
\hat{g}_{\mu \nu} \nonumber \\
g_{ij} &=& e^{2/3 \phi} \hat{g}_{ij} = \hat{g}_{zz}^{-1/2}(
\hat{g}_{ij}\hat{g}_{zz} - \hat{g}_{iz} \hat{g}_{jz} ), \eea where
we have used $e^{4/3 \phi} = \hat{g}_{zz}$ and $A_i =
\hat{g}_{iz}/\hat{g}_{zz}$.

The ansatz for the reduction of the 3-form field $\hat{C}_3$ in
eleven dimensions is as follows \be \label{ansatzB}
\hat{C}_3(x^{\mu},z) = C_3(x^{\mu}) + B(x^{\mu}) \wedge dz. \ee
Differentiating both sides we find the ansatz for the reduction of
the eleven dimensional 4-form field strength $\hat{F}_4 =
d\hat{C}_3$. A convenient way of writing it is (see e.g.
\cite{julia}) \be \label{ansatzF} \hat{F}_4 = F_4 + F_3 \wedge (dz
+ A), \ee where \be \label{ansatzF2} F_4 = dC_3 - dB \wedge A, \ \
\ \ \ F_3 = dB. \ee

After dimensional reduction to $D=10$, the next step is to perform
the deformation along two of the isometry coordinates which we
label  as $x^1$ and $x^2$ as above. Then the 3-torus that we use
to generate a new solution has coordinates $\{x^1,x^2,x^5\}$.
Because of our first assumption we have $g_{\mu i} = 0$ for all
$i=1,\cdots,4$ and $ \mu = 6, \cdots, 11$. Moreover, looking at
the ansatz (\ref{ansatzF2}) we see that our assumption on ${\hat
F}_4$ implies that $B_{ij}= B_{i\mu}=0$. Thus we have $d=4$ in
(\ref{matrix}). \footnote{It might happen that one or two of the
remaining isometries $\{x^3,x^4\}$ have no couplings with $x^1$
and $x^2$, in which case it would be enough to choose $d=2$ or
$d=3$, respectively. However, we prefer to consider the most
general case with $d=4$, as the others can be studied within this
formalism.} Also note that background matrix (\ref{background})
$E$ symmetric, that is $E=g_{ij}$. From (\ref{onemli}),
(\ref{matrix}), (\ref{matrixc}) and (\ref{dilaton}) we get
\begin{eqnarray}\label{100}
\tilde{g}_{\mu \nu}& = &g_{\mu \nu}, \ \ \ \mu, \nu = 6,\cdots,11 \nonumber  \\
\tilde{g}_{\mu i} & = & g_{\mu i} = 0, \ \ \ \ \ i = 1,2,3,4
\nonumber \\
\tilde{g}_{wi}& = &G g_{wi}, \ \ \ \ w = 1,2  \nonumber \\
\tilde{g}_{34}& = &G g_{34} +  G \gamma^2 detg(3 \mid 4)
\nonumber\\
\tilde{g}_{33}& = &G g_{33} +  G \gamma^2 detg(4 \mid 4) \nonumber
\\
\tilde{g}_{44} &=& G g_{44} +  G \gamma^2 detg(3 \mid 3) \\
\nonumber \\
\tilde{B}_{\mu \nu} &=& B_{\mu \nu}, \ \ \ \mu, \nu = 6,\cdots,11 \nonumber  \\
\label{compactB}
\tilde{B}_{\mu i} & = & B_{\mu i} = 0, \ \ \ \ i=1,2,3,4 \\
\tilde{B}_{ij} &=& \frac{1}{2} G \gamma \epsilon^{ij} \sum_{k,l
\neq i,j ;\  k \neq l} detg(k,l \mid 3,4), \ \ \  i,j,k,l \in
\{1,2,3,4\} \nonumber \\
\nonumber \\
\label{102} e^{2 \tilde{\phi} } &=&  G e^{2 \phi}
\end{eqnarray}
where we have defined \be \label{G} G ={\rm det}(\Gamma_d g +
1_d)^{-1} = [1 + \gamma^2 detg(3,4 \mid 3,4)]^{-1}. \ee Here $g$
is the $4 \times 4$ matrix with entries $g_{ij}$ and our
convention for the B-field is such that \be B = \frac{1}{2}B_{IJ}
dx^I \wedge dx^J \, , \ \ \ \ \  I,J=1, ...,10 \, , \ee and
$\epsilon^{IJ}$ is an antisymmetric tensor with $\epsilon^{IJ} =
1$ for $I < J$. As can be seen from (\ref{ansatzB}), the lifting
of $\tilde{B}_{IJ}$ to $D=11$ brings the  contribution $\tilde{B}
\wedge dz$ to the 3-form field in the deformed $D=11$ background,
which is linear in the deformation parameter $\gamma$.  An extra
contribution comes from the lifting of the deformed 3-form field
$C_3$ in $D=10$. Remember that because of our condition on ${\hat
F}_4$ we have  $B_{ij}=B_{i\mu}=0$. Also note that  $F_2 = dA =
d(A_i(x^{\mu})dx^i)$ can have at most one component along the
isometry coordinates. Then all contraction terms in (\ref{f2f4})
are zero except the $\gamma i_1 i_2 F_6 = \gamma i_1 i_2
\star_{10} F_4$ term. Thus, \bea
\label{defF2} \tilde{F}_2 & = & F_2  \\
\tilde{F}_4 & = & \label{deformedF10} F_4 - F_2 \wedge B - \gamma
i_{1} i_{2} \star_{10} F_4 + F_2 \wedge \tilde{B}   =  F_4 -
\gamma i_{1} i_{2} \star_{10} F_4 + F_2 \wedge \tilde{B}^{||} \eea
\noindent where in the last line we have decomposed the deformed
B-field into its part $\tilde{B}^{||}$ along the isometry
directions and the part $\tilde{B}^{\bot}$ transverse to the
isometry directions \be \tilde{B} = \tilde{B}^{||} +
\tilde{B}^{\bot} = \tilde{B}^{||} + B \, . \ee

\noindent Substituting $\tilde{F}_4$ and $\tilde{B}$ in
(\ref{ansatzF}), we find the deformed 4-form field
$\tilde{\hat{F}}_4$ in eleven dimensions: \bea \label{deformedF11}
\tilde{\hat{F}}_4 &=& F_4 - \gamma i_{1} i_{2} \star_{10}
F_4 + F_2 \wedge \tilde{B}^{||} + d(\tilde{B}^{||} +B) \wedge (dz + \tilde{A}) \nonumber \\
& = &\hat{F}_4 - \gamma i_{1} i_{2} \star_{10} F_4 +
d[\tilde{B}^{||} \wedge (dz + A)],  \eea where the Hodge star is
taken in ten dimensions with respect to the undeformed metric. We
also use the fact that $\tilde{A} = A$, as it follows from
(\ref{defF2}). This formula can be written in terms of the eleven
dimensional fields as (see Appendix \ref{A} for the proof): \be
\label{deformedF11asil} \tilde{\hat{F}}_4 = \hat{F}_4 - \gamma
i_{1} i_{2} i_{z} \star_{11} \hat{F}_4 + \frac{\gamma}{2} d\left(G
\sum_{q, r \neq m,n,p ; \ q \neq r} det\hat{g}(q,r \mid
3,4)\frac{\epsilon^{mnp}dx_m \wedge dx_n \wedge dx_p}{3!}\right),
\ee where $m,n,p,q,r \in \{1,2,3,4,z=5 \}$ and $\hat{g}$ is the $5
\times 5$ torus-matrix with entries $ \hat{g}_{mn}$ and we used
$\{x^1,x^2,x^5\}$ directions for the deformation. The Hodge dual
$\star_{11}$ is taken in the 11-dimensional space, with respect to
the undeformed metric. Also $\epsilon^{mnp} = 1$ when $m < n < p$
and it is totally antisymmetric in its indices.

Now in order to lift the ten dimensional deformed metric back to
eleven dimensions we have to substitute (\ref{100}) in the
reduction ansatz (\ref{ansatz}). As a result we have \be
\label{metric11} \tilde{ds}_{11}^2 = e^{-2/3 \tilde{\phi}}
\tilde{g}_{\mu \nu} dx^{\mu} dx^{\nu} + e^{4/3 \tilde{\phi}} (dz +
A)^2, \ee where we have used $\tilde{A} = A$. Using (\ref{102})
and (\ref{100}) together with (\ref{metric11a}), (\ref{hatbar})
and the fact that $e^{4/3 \phi} = \hat{g}_{zz}$, we can write the
deformed eleven dimensional metric (\ref{metric11}) as:
\begin{eqnarray} \label{metric111} \tilde{ds}_{11}^2 &=& G^{-1/3}
e^{-2/3 \phi} [G g_{ij}dx^{i} dx^{j}  + g_{\mu \nu}dx^{\mu}
dx^{\nu} + G\gamma^2 (detg(4 \mid 4) dx^{3} dx^{3}
\nonumber \\
&& +   detg(3 \mid 4) dx^{3} dx^{4} + detg(3 \mid 3) dx^{4}
dx^{4})] + G^{2/3} e^{4/3 \phi}
(dz + A)^2 \nonumber \\
& = & G^{-1/3}\hat{g}_{\mu \nu} dx^{\mu} dx^{ \nu}  + G^{2/3}
\hat{g}_{mn} dx^m dx^n  \\
&& + G^{2/3} \gamma^2 \hat{g}_{zz}^{-1/2} [detg(4 \mid 4) dx^{3}
dx^{3} + detg(3 \mid 4) dx^{3} dx^{4} + detg(3 \mid 3) dx^{4}
dx^{4}],\nonumber
\end{eqnarray}
Now using the following facts (see Appendix \ref{B} for the proof)
\bea \label{equality} && \hat{g}_{zz}^{-1/2} [detg(i \mid j)] =
det\hat{g}(i \mid j), \nonumber \\ && detg(i,j \mid k,l) =
det\hat{g}(i,j \mid k,l), \ \ \ i,j = 1,2,3,4 \eea we can write
the metric in (\ref{metric111}) completely in terms of the eleven
dimensional fields: \bea \label{esassonuc} d\tilde{s}_{11}^2 &=&
G^{-1/3}\hat{g}_{\mu \nu} dx^{\mu} dx^{ \nu} + G^{2/3}
\hat{g}_{mn} dx^m dx^n  \\ && + G^{2/3} \gamma^2 [det\hat{g}(4
\mid 4) dx^{3} dx^{3} + det\hat{g}(3 \mid 4) dx^{3} dx^{4} +
det\hat{g}(3 \mid 3) dx^{4} dx^{4}]. \nonumber \eea Using
(\ref{equality}) in (\ref{G}), $G$ can be expressed as \be
\label{hatG} G = [1 + \gamma^2 det\hat{g}(3,4 \mid 3,4)]^{-1} \, .
\ee As we see, our new solution is expressed in terms of the
original one and subdeterminants of the torus matrix $\hat{g}$.
Hence, there is no need anymore to refer to $D=10$ or details of
the derivation. Note that we have additive correction terms to
$ds_{11}^2$ only along the isometry coordinates, which are not
involved in the deformation process namely along $x^3$ and $x^4$.
Also the correction terms $det\hat{g}(r \mid s), r,s = 3,4$ are
invariant under the relabeling of the indices $\{1,2,5\}$. If we
interchange the indices $1 \leftrightarrow 2, \ 1 \leftrightarrow
5$, or $2 \leftrightarrow 5$ in the matrix $\hat{g}(r \mid s)$,
the resulting matrix will have the same determinant as one can
pass from one to the other by equal number of row and column
interchanges. It is also easy to see that the correction terms
coming to the deformed 4-form field $\tilde{\hat{F_4}}$ in
(\ref{deformedF11asil}) are invariant under the relabeling of the
indices $1 \leftrightarrow 2, \ 1 \leftrightarrow 5$, or $2
\leftrightarrow 5$.\footnote{More precisely, $\tilde{\hat{F}}_4$
will remain the same under cyclic permutations of these three
coordinates but $\gamma$ terms will pick up an overall -1 sign
otherwise. However, this sign change can be eliminated by changing
the orientation of the 5-torus or sending $\gamma \to -\gamma$.}
Consequently, once we fix the deformation 3-torus with coordinates
$\{x^1, x^2, z = x^5\}$, it does not make a difference as to how
we choose the reduction and the T-duality coordinates; the
resulting deformed metric $d\tilde{s}_{11}^2$ and the deformed
4-form field $\tilde{\hat{F_4}}$  will always be the same.

In deriving our main equations (\ref{deformedF11asil}),
(\ref{esassonuc}) and (\ref{hatG}), we assumed the existence of
five  $U(1)$ isometries. However, let us emphasize that these
formulas are also valid if $x^3$ and $x^4$ are not $U(1)$
directions but just some coordinates that mix with the deformation
3-torus $\{x^1,x^2,x^5\}$ in the metric. If there are no such
couplings or if the the original background has only three or four
$U(1)$ isometries, then our formulas can easily be modified. The
case when the background has only four commuting isometries  that
decouples from the rest can be regarded as a special case for our
general method, in which $x^4$ (or $x^3$) does not mix with the
remaining isometry coordinates. When $x^4$ does not mix,  we have
$det\hat{g}(3 \mid 3)= \hat{g}_{44} det\hat{g}(3, 4 \mid 3,4)$, as
a result of which the coefficient of $dx^4dx^4$ in the metric
(\ref{esassonuc}) becomes $G^{-1/3}\hat{g}_{44}$. Also
$det\hat{g}(3 \mid 4)$ vanishes, so that the only additional term
in the deformed metric is to the term $dx^3dx^3$. This happens
frequently in the examples below, for instance in $\beta$
deformations when $x^4$ is the $AdS_4$ isometry direction. Now the
general formulas become: \bea \tilde{\hat{F}}_4 &=& \hat{F}_4 -
\gamma i_{1} i_{2} i_{z} \star_{11} \hat{F}_4 + \gamma d\left(G
\sum_{q \neq m,n,p} det\hat{g}(q \mid 3)\frac{\epsilon^{mnp}dx_m
\wedge dx_n
\wedge dx_p}{3!} \right), \nonumber \\
\label{4U1} d\tilde{s}_{11}^2 &=& G^{-1/3}\hat{g}_{\mu \nu}
dx^{\mu} dx^{ \nu} + G^{2/3} \hat{g}_{mn} dx^m dx^n + G^{2/3}\gamma^2 det \hat{g} \, dx^3dx^3 \, , \\
G &=& [1 + \gamma^2 det\hat{g}(3 \mid 3)]^{-1} \, , \nonumber \eea
where $\hat{g}$ is the $4 \times 4$ matrix with entries $
\hat{g}_{mn}, \ m,n = \{1,2,3,5\}$. Here $x^3$ is not necessarily
a $U(1)$ direction, and if so this would be the result with $n=3$
where the deformation 3-torus directions have couplings with
$x^3$. Similarly, backgrounds with 3 commuting decoupled
isometries can be regarded as a special case, for which both $x^3$
and $x^4$ do not mix with the 3 remaining isometry coordinates.
Let $\hat{g}$ denote the  $3 \times 3$ torus matrix that
corresponds to the remaining $U(1)$ directions $\{x^1,x^2,x^5\}$.
Then, we have \bea \tilde{\hat{F}}_4 &=& \hat{F}_4 - \gamma i_{1}
i_{2} i_{z} \star_{11} \hat{F}_4 + \gamma d\left(G det\hat{g} \,
dx_1 \wedge dx_2 \wedge dx_5
\right), \nonumber \\
\label{3U1} d\tilde{s}_{11}^2 &=& G^{-1/3}\hat{g}_{\mu \nu}
dx^{\mu} dx^{ \nu} + G^{2/3} \hat{g}_{mn} dx^m dx^n \, , \\
G &=& [1 + \gamma^2 det\hat{g}]^{-1} \, ,\nonumber \eea where $m,
n$ =$\{1,2, 5\}$.

Now we  apply our results to some examples. In some of them (such
as the one in the next subsection) one or two of the unused
isometry directions $\{x^3,x^4\}$ have no coupling with the
deformation directions $\{x^1, x^2, x^5\}$. As we have just
explained above, in such cases  it is possible to work with $4
\times 4$ or $3 \times 3$ torus matrices. However, to make our
presentation more coherent we will always take the background
matrices as $5 \times 5$.

\subsection{Example 1: $AdS_4 \times (\textrm{Sasaki-Einstein})_7$ ({\rm with base} $CP^2$)}
In this subsection we will consider the $\beta$ and dipole
deformations of the background
$$AdS_4 \times Y_7$$ where $Y_7$ is the seven dimensional
Sasaki-Einstein space found recently by \cite{sasaki2}  with base
$CP^2$. Although, its $\beta$ deformation was already
obtained in \cite{jerome}, we will begin with that example to
illustrate our method.

For this background our 11-dimensional metric and the 4-form field
are \be \label{cp2} ds^2_{11}= ds^2_{AdS_4} + ds^2_{Y_7} \, ,
\hspace{1cm} \hat{F}_4= 6 \, vol(AdS_4) \ee where $AdS_4$ and
$Y_7$ metrics are given after suitable scalings as
\be
ds^2_{AdS_4} = -(1 +  r^2)dt^2 + \frac{dr^2}{1 + r^2} + r^2
(d\chi_1^2 + \sin^2 \chi_1 d\chi_2^2) \,\, , \label{adsmetric}
\ee
and
\bea
ds^2_{Y_7} &=& U^{-1}d\rho^2 + 3 \rho^2\left(\mu_1^2d\phi_1^2 + \mu_2^2d\phi_2^2 - [\mu_1^2d\phi_1 + \mu_2^2d\phi_2]^2 + \sum_{i=1}^3 d\mu_i^2 \right) \nonumber \\
&& + q(d\psi + j_1)^2 + \o [d\alpha + f(d\psi + j_1)]^2 \, .
\eea
Here $U, q, \o, f$ are some functions of $\rho$ (for details see
\cite{jerome}), $j_1= 3(\mu_1^2d\phi_1 +\mu_2^2d\phi_2)$ and
$\sum_{i=1}^3\mu_i^2=1$. The $U(1)$ isometries of this background
correspond to Killing vectors $(\partial_{\phi_1},
\partial_{\phi_2}, \partial_{\chi_2}, \partial_{\alpha}, \partial_{\psi})$, where the last one is the R-symmetry direction.\footnote{$AdS_4$ has a
further isometry which corresponds to the shift of the time
coordinate. However, we are not going to use this isometry.}

\subsubsection{$\beta$ Deformations}
Let us label the 5-torus directions $(\partial_{\phi_1},
\partial_{\phi_2}, \partial_{\psi}, \partial_{\chi_2}, \partial_{\alpha})$
as $\{x^1,..., x^5\}$ respectively. That means that we choose our
3-torus for the deformation as $(\phi_1, \phi_2, \alpha)$ and
therefore avoid using the R-symmetry $\partial_{\psi}$ in the
deformation process. The reduction direction is $z=\alpha$. Then,
the 5-torus matrix is \be \label{bgcp2} \hat{g}=
\left(\begin{array}{ccccc} 9 \mu_1^4\delta +
3\rho^2\mu_1^2(1-\mu_1^2)& 3\mu_1^2\mu_2^2(3\delta-\rho^2)
& 3\mu_1^2 \delta& 0 & 3f \o\mu_1^2 \\
. & 9 \mu_2^4\delta + 3\rho^2\mu_2^2(1-\mu_2^2) \, &3\mu_2^2 \delta & 0 & 3f\o\mu_2^2 \\
. & . & \delta & 0 & \o f \\
. & . & . & r^2 \sin^2\chi_1 & 0 \\
. & . & . & . & \o
 \end{array}\right) \, ,
\ee where $\delta \equiv (q+ \o f^2)$ and $\hat{g}$ is a symmetric
matrix. The nonzero subdeterminants are\bea
det \hat{g}(4\mid 4) &=& 9 \rho^4q \o \mu_1^2\mu_2^2\mu_3^2 \, , \nonumber \\
det \hat{g}(3\mid 3) &=& 9\mu_1^2\mu_2^2\rho^2 \o
r^2\sin^2\chi_1[\mu_3^2\rho^2+3q(\mu_1^2+\mu_2^2)] \, ,
\nonumber \\
det \hat{g}(3,4 \mid 3,4) &=& 9 \mu_1^2 \mu_2^2 \rho^2\o [\mu_3^2\rho^2 + 3q(\mu_1^2+\mu_2^2)] \,, \\
det \hat{g}(4,5 \mid 3,4) &=& 9 \mu_1^2\mu_2^2\mu_3^2\rho^4\o f \,
,
\nonumber \\
det \hat{g} (1,4 \mid 3,4) &=& -9 \mu_1^2\mu_2^2\rho^2\o q = - det
\hat{g}(2,4 \mid 3,4) \, . \nonumber \eea Then, the eleven
dimensional deformed metric from (\ref{esassonuc}) is \bea
\label{cp2beta} d\tilde{s}_{11}^2 &=& G^{-1/3}[ds^2_{AdS_4}+
U^{-1}d\rho^2 +
3\rho^2\sum_{i=1}^3 d\mu_i^2] + G^{2/3} \{\g^2 9\mu_1^2\mu_2^2\mu_3^2 \rho^4 \o q d\psi^2  \\
&+&  3 \rho^2\left(\mu_1^2d\phi_1^2 + \mu_2^2d\phi_2^2 -
[\mu_1^2d\phi_1 + \mu_2^2d\phi_2]^2  \right)+ q(d\psi + j_1)^2 +
\o [d\alpha + f(d\psi + j_1)]^2\} \nonumber \eea where \be G^{-1}
= 1 + \g^2 \left(9 \mu_1^2 \mu_2^2 \rho^2\o [\mu_3^2\rho^2 +
3q(\mu_1^2+\mu_2^2)]\right). \ee In writing the metric we used the
fact that $det\hat{g}(3 \mid 3) = g_{44} det\hat{g}(3,4 \mid
3,4)$.

The Hodge dual of the 4-form field is \be \star_{11}\hat{F}_4 = 54
\rho^4 (\frac{q \o}{U})^{1/2} \mu_1 \mu_2 \sqrt{1+\mu_1^2+\mu_2^2}
\, d\rho \wedge d\phi_1 \wedge d\phi_2 \wedge d\mu_1 \wedge d\mu_2
\wedge d\psi \wedge d\alpha \ee Then using (\ref{deformedF11asil})
we find the deformed 4-form field $\tilde{\hat{F}}_4$ as \bea
\label{sasakifcp2baz} \tilde{\hat{F}}_4 & = & \hat{F}_4 - 54 \g
\rho^4 (\frac{q \o}{U})^{1/2} \mu_1 \mu_2 \sqrt{1+\mu_1^2+\mu_2^2}
\, d\rho \wedge d\mu_1 \wedge d\mu_2 \wedge d\psi \nonumber \\ &&
+ \g d\{9 G  \mu_1^2 \mu_2^2 \rho^2 \o [(\mu_3^2\rho^2 +
3q(\mu_1^2+\mu_2^2))d\phi_1 \wedge d\phi_2 \wedge d\alpha + q
(d\phi_1 - d\phi_2) \wedge d\psi \wedge d\alpha \nonumber \\ &&
 + \mu_3^2 \rho^2 f d\phi_1 \wedge d\phi_2 \wedge d\psi]\}. \eea
These agree with the results of \cite{jerome} and here we have the
additive correction term in the metric (\ref{cp2beta}), that is $
G^{2/3}\g^2 9\mu_1^2\mu_2^2\mu_3^2 \rho^4 \o q d\psi^2$, written
explicitly.

\subsubsection{Dipole Deformations}
Now we apply our method to obtain dipole deformations of the above
background (\ref{cp2}). The necessary torus matrix is obtained
from (\ref{bgcp2}) by interchanging its rows and columns in the
order $4 \leftrightarrow 5, 3 \leftrightarrow 4$ which gives the
following symmetric matrix \be \hat{g}= \left(\begin{array}{ccccc}
9 \mu_1^4\delta + 3\rho^2\mu_1^2(1-\mu_1^2)&
3\mu_1^2\mu_2^2(3\delta-\rho^2)
&  3f\o \mu_1^2 \,\, & 3\mu_1^2 \delta & 0  \\
. & 9 \mu_2^4\delta + 3\rho^2\mu_2^2(1-\mu_2^2) \,\, & 3f\o \mu_2^2 \, &3\mu_2^2 \delta & 0  \\
. & . & \o & \o f & 0 \\
. & . & . & \delta &  0 \\
. & . & . & . & r^2\sin ^2\chi_1
 \end{array}\right) \, .
\label{bgcp2dipole} \ee Here $x^1=\phi_1, x^2=\phi_2, x^3=\alpha,
x^4=\psi, x^5=\chi_2$. So, the deformation 3-torus is
$\{\phi_1,\phi_2,\chi_2\}$ and again we don't use the R-symmetry
$\partial_{\psi}$ . The relevant nonzero subdeterminants are \bea
det \hat{g} (4 \mid 4) &=& 9\mu_1^2\mu_2^2\rho^2
\o r^2\sin^2\chi_1[\mu_3^2\rho^2+3q(\mu_1^2+\mu_2^2)] \, , \nonumber \\
det \hat{g} (3 \mid 4) &=& 9 \mu_1^2\mu_2^2\mu_3^2\rho^4\o f r^2 \sin^2 \chi_1 \, , \nonumber\\
det \hat{g} (3 \mid 3) &=& 9 \mu_1^2\mu_2^2\mu_3^2\rho^4 (q + \o f^2) r^2 \sin^2 \chi_1 \, , \nonumber\\
det \hat{g} (3, 4 \mid 3, 4) &=& 9 \mu_1^2\mu_2^2 \rho^2 [ 3 (q + \o f^2)(\mu_1^2+\mu_2^2) + \rho^2\mu_3^2] r^2 \sin^2 \chi_1 \, , \\
det \hat{g} (1, 4 \mid 3, 4) &=& -9 \mu_1^2\mu_2^2 \rho^2 \o f r^2 \sin^2 \chi_1 = -det \hat{g} (2, 4 \mid 3, 4) \, , \nonumber\\
det \hat{g} (1, 3 \mid 3, 4) &=&  -9 \mu_1^2\mu_2^2 \rho^2  (q +
\o f^2) r^2 \sin^2 \chi_1 = -det \hat{g} (2, 3 \mid 3, 4) \, .
\nonumber \eea Then the deformed metric from (\ref{esassonuc}) is
\bea d\tilde{s}_{11}^2 &=& G^{-1/3}[ds^2_{AdS_4} - r^2\sin^2\chi_1
d\chi_2^2 + U^{-1}d\rho^2 +
3\rho^2\sum_{i=1}^3 d\mu_i^2]  + G^{2/3}[r^2\sin^2\chi_1 d\chi_2^2 \nonumber\\
&+&  3 \rho^2\left(\mu_1^2d\phi_1^2 + \mu_2^2d\phi_2^2 -
[\mu_1^2d\phi_1 + \mu_2^2d\phi_2]^2  \right)+ q(d\psi + j_1)^2 +
\o [d\alpha + f(d\psi + j_1)]^2] \nonumber \\
&+&  G^{2/3} \g^2 9\mu_1^2\mu_2^2\rho^2 r^2
\sin^2\chi_1[(\mu_3^2\rho^2 +3q(\mu_1^2+\mu_2^2))\o d\alpha^2 +
\mu_3^2\rho^2\o f d\alpha d\psi \nonumber \\
&+& \mu_3^2\rho^2 (q + \o f^2)d\psi^2] \eea where \be G^{-1} = 1 +
\g^2 \left(9 \mu_1^2\mu_2^2 \rho^2 [ 3 (q + \o
f^2)(\mu_1^2+\mu_2^2) + \rho^2\mu_3^2] r^2 \sin^2 \chi_1\right).
\ee

Using (\ref{deformedF11asil}) we find the deformed 4-form field
$\tilde{\hat{F}}_4$ as \bea \label{sasakifcp2bazdipole}
\tilde{\hat{F}}_4 & = & \hat{F}_4 + \g d\{9 G \mu_1^2\mu_2^2
\rho^2 r^2 \sin^2 \chi_1 [( 3 (q + \o f^2)(\mu_1^2+\mu_2^2) +
\rho^2\mu_3^2) d\phi_1 \wedge d\phi_2 \wedge d\chi_2 \nonumber \\
&&+ \o f(d\phi_1 - d\phi_2) \wedge d\alpha \wedge d\chi_2 + (q +
\o f^2)(d\phi_1 - d\phi_2)\wedge d\psi \wedge d\chi_2]\}. \eea

\subsection{Example 2: $AdS_7 \times S^4$}
Our next background is $AdS_7 \times S^4$. Noncommutative and
dipole  deformations of the $M5$-brane solution were studied in
\cite{berman}, where  the near horizon limits was also explained.
Here we directly start from the $AdS_7 \times S^4$ geometry and
use a metric parametrization different from \cite{berman}. This
background can be written as (after suitable rescalings): \be
\label{ads7} ds^2=-(1+r^2)dt^2 + \frac{dr^2}{1+r^2} +
r^2d\Omega_5^2 + d\Omega_4 \ee where spheres are parametrized as
\bea \label{sphere} && d\Omega_5^2 = d\alpha^2 +
c_{\alpha}^2(d\phi_3 -d\phi_2)^2 + s_{\alpha}^2 [d\theta^2 +
\cos^2\theta(d\phi_3 + d\phi_1+d\phi_2)^2
+ \sin^2\theta(d\phi_3-d\phi_1)^2]  \nonumber \\
&& d\Omega_4^2 = d\chi_1^2 +\cos^2\chi_1 d\chi_2^2+
\sin^2\chi_1[d\chi_3^2 + \sin^2\chi_3 d\chi_4^2] \eea $U(1)$
directions are $\{\phi_1, \phi_2, \phi_3\}$ and $\{\chi_2,
\chi_4\}$. The 4-form field strength is given as \be
\label{f4dipole} \hat{F}_4 = vol(\Omega_4) = \cos\chi_1
\sin^2\chi_1 \sin\chi_3 d\chi_1 \wedge d\chi_2 \wedge d\chi_3
\wedge d\chi_4. \ee In finding the deformed 4-form field in eleven
dimensions we use the following orientation \be
\star_{11}\hat{F}_4 = vol(AdS_7) = \frac{r^5 \cos \alpha \sin^3
\alpha \cos \theta \sin \theta}{3} dt \wedge dr \wedge d\alpha
\wedge d\theta \wedge d\phi_1 \wedge d\phi_2 \wedge d\phi_3. \ee

\subsubsection{Noncommutative Deformations}

We perform a noncommutative deformation by choosing all the
isometries from the $AdS_7$ part: $x^1 = \phi_1, x^2 = \phi_2, x^5
= z = \phi_3$. We also label $x^3 = \chi_2, x^4 = \chi_4$. Then
the torus matrix is ($s_{\alpha} \equiv \sin \alpha, \,\,
c_{\alpha} \equiv \cos \alpha$ and $s_{\theta} \equiv \sin \theta,
\,\, c_{\theta} \equiv \cos \theta$) \be \hat{g} =
\left(\begin{array}{ccccc}
r^2s_{\alpha}^2 & r^2s_{\alpha}^2  \cos^2 \theta  & 0 & 0 & r^2 s_{\alpha}^2 \cos2\theta \\
. & r^2(c_{\alpha}^2+s_{\alpha}^2\cos^2\theta)  & 0 & 0 &
r^2(-c_{\alpha}^2+s_{\alpha}^2\cos^2\theta)
\\ . & .   & \cos^2\chi_1 & 0 & 0 \\
. & . & . & \sin^2\chi_1\sin^2 \chi_3  & 0\\
. & . & . & . & r^2
\end{array}\right) \ee
which is symmetric. The relevant nonzero subdeterminants are \bea
det \hat{g} (3, 4 \mid 3, 4) &=& 9 r^6 c_{\alpha}^2 s_{\alpha}^4
s_{\theta}^2 c_{\theta}^2 \equiv \Delta \, , \nonumber\\ det
\hat{g} (4 \mid 4)
&=& \Delta \cos^2\chi_1 \, , \\
det \hat{g} (3 \mid 3) &=& \Delta \sin^2\chi_1 \sin^2\chi_3 \, ,
\nonumber \eea The deformed metric is found from (\ref{esassonuc})
as \bea d\tilde{s}_{11}^2 &=& G^{-1/3}[-(1+r^2)dt^2 +
\frac{dr^2}{1+r^2} + r^2 (d\alpha^2 + \sin^2 \alpha \, d\theta^2)
+
d\Omega_4]  \\
&+& G^{2/3}[\cos^2\alpha(d\phi_3 -d\phi_2)^2 + \sin^2 \alpha (
\cos^2\theta(d\phi_3 + d\phi_1+d\phi_2)^2 +
\sin^2\theta(d\phi_3-d\phi_1)^2)], \nonumber  \eea where \be
G^{-1} = 1 + \g^2 \Delta. \ee Note that having $det\hat{g}(3 \mid
3)= \hat{g}_{44} det\hat{g}(3, 4 \mid 3,4)$ and $det\hat{g}(4 \mid
4)= \hat{g}_{33} det\hat{g}(3, 4 \mid 3,4)$ has simplified our
results, as we discussed earlier. On the other hand, using
(\ref{deformedF11asil}) the deformed 4-form field is found to be
\bea \label{ncf} \tilde{\hat{F}}_4 & = & \hat{F}_4 + \g  \frac{r^5
c_{\alpha} s_{\alpha}^3 c_{\theta} s_{\theta}}{3} dt \wedge dr
\wedge d\alpha \wedge d\theta + \g d \left( \frac{\Delta}{1 + \g^2
\Delta} d\phi_1 \wedge d\phi_2 \wedge d\phi_3 \right). \eea

\subsubsection{Type I Dipole Deformations}
As noted in \cite{berman} there are two different ways to do the
dipole deformations of $AdS_7\times S^4$. The first option is to
choose two isometries from the $AdS$ part and one isometry from
the $S^4$ part and the second option is to choose two isometries
from the $S^4$ and one from the $AdS_7$. Following the terminology
introduced in \cite{berman} we call them Type I and Type II
deformations,  respectively. In this subsection we will consider
only the Type I case and leave discussion of the other to the
subsection \ref{exceptional}. The Type I case is in agreement with
our assumption on $\hat{F}_4$ and hence we can use our general
formulas (\ref{deformedF11}) and (\ref{esassonuc}). For example,
labeling directions as $x^1= \phi_1, x^2=\phi_2, x^3=\phi_3,
x^4=\chi_2, x^5=\chi_4$ the torus matrix is  ($s_{\alpha} \equiv
\sin \alpha, \,\, c_{\alpha} \equiv \cos \alpha$ and $s_{\theta}
\equiv \sin \theta, \,\, c_{\theta} \equiv \cos \theta$): \be
\hat{g} = \left(\begin{array}{ccccc}
r^2s_{\alpha}^2 & r^2s_{\alpha}^2 \cos^2 \theta & r^2 s_{\alpha}^2 \cos2\theta & 0 & 0 \\
. & r^2(c_{\alpha}^2+s_{\alpha}^2\cos^2\theta) &
r^2(-c_{\alpha}^2+s_{\alpha}^2\cos^2\theta) & 0 & 0
\\ . & . & r^2 & 0 & 0 \\
. & . & . & \cos^2\chi_1  & 0\\
. & . & . & . & \sin^2\chi_1\sin^2 \chi_3
\end{array}\right). \ee
The relevant nonzero subdeterminants are \bea det \hat{g} (4 \mid
4) &=& 9 r^6 c_{\alpha}^2 s_{\alpha}^4 s_{\theta}^2 c_{\theta}^2 \sin^2\chi_1 \sin^2\chi_3 \, , \nonumber \\
det \hat{g} (3 \mid 3) &=& r^4 \sin^2\chi_1 \sin^2\chi_3
\cos^2\chi_1 s_{\alpha}^2(c_{\alpha}^2 +
s_{\alpha}^2 c_{\theta}^2 s_{\theta}^2) \, , \nonumber\\
det \hat{g} (3, 4 \mid 3, 4) &=& r^4 \sin^2\chi_1 \sin^2\chi_3
s_{\alpha}^2(c_{\alpha}^2 +
s_{\alpha}^2 c_{\theta}^2 s_{\theta}^2) \equiv \Delta \, , \nonumber\\
det \hat{g} (1, 4 \mid 3, 4) &=& r^4 \sin^2\chi_1 \sin^2\chi_3
s_{\alpha}^2[-c_{\alpha}^2 (c_{\theta}^2 + c_{2\theta})+
s_{\alpha}^2 c_{\theta}^2 s_{\theta}^2] \, , \nonumber\\
det \hat{g} (2, 4 \mid 3, 4) &=& r^4 \sin^2\chi_1 \sin^2\chi_3
s_{\alpha}^2(-c_{\alpha}^2 + 2s_{\alpha}^2 c_{\theta}^2
s_{\theta}^2).
 \eea
Then the deformed metric using (\ref{esassonuc}) is \bea
d\tilde{s}_{11}^2 &=& G^{-1/3}[-(1+r^2)dt^2 + \frac{dr^2}{1+r^2} +
r^2 (d\alpha^2 + \sin^2 \alpha \, d\theta^2) +
d\Omega_4]  \\
&& + G^{2/3}[\cos^2\alpha(d\phi_3 -d\phi_2)^2 + \sin^2 \alpha (
\cos^2\theta(d\phi_3 + d\phi_1+d\phi_2)^2 +
\sin^2\theta(d\phi_3-d\phi_1)^2)] \nonumber \\
&& + G^{2/3} \g^2 \sin^2\chi_1 \sin^2\chi_3[\Delta \, d\chi_4^2 +
9 r^6 c_{\alpha}^2 s_{\alpha}^4 s_{\theta}^2 c_{\theta}^2 \,
d\phi_3^2],
 \eea where \be
G^{-1} = 1 + \g^2 \Delta. \ee Again we have used the fact that
$det\hat{g}(3 \mid 3)= \hat{g}_{44} det\hat{g}(3, 4 \mid 3,4)$ and
the identity $G^{2/3} - G^{-1/3} = G^{2/3} \g^2 \Delta$.
Meanwhile, the deformed 4-form field from (\ref{deformedF11asil})
is \bea \label{s4dipolef} \tilde{\hat{F}}_4 & = & \hat{F}_4  + \g
d [G r^4 \sin^2\chi_1 \sin^2\chi_3 s_{\alpha}^2((c_{\alpha}^2 +
s_{\alpha}^2 c_{\theta}^2 s_{\theta}^2)) d\phi_1 \wedge d\phi_2
\wedge d\chi_4  \\ && + ((-c_{\alpha}^2 (c_{\theta}^2 +
c_{2\theta})+ s_{\alpha}^2 c_{\theta}^2 s_{\theta}^2)) d\phi_2
\wedge d\phi_3 \wedge d\chi_4 + ((-c_{\alpha}^2 + 2s_{\alpha}^2
c_{\theta}^2 s_{\theta}^2)) d\phi_1 \wedge d\phi_3 \wedge
d\chi_4]. \nonumber \eea

\subsection{An Exceptional Case: Type II Dipole Deformations of $AdS_7 \times S^4$}
\label{exceptional}

Until now we have discussed 1-parameter deformations, with the
assumption that ${\hat F}_4$ has at most one leg along the
isometry directions. In this section, we illustrate how our method
works, when this assumption does not hold. We  do this by
considering the dipole deformation of  the $AdS_7 \times S^4$
example (\ref{ads7}) where the 3-torus associated with  the
deformation has coordinates $\{\phi_1, \chi_2, \chi_4\}$. At first
sight, there seems to be two options to perform the deformation,
which might yield different results. One possibility is to start
by reducing along an $S^4$ coordinate ($\chi_2 $ or $\chi_4$).
This gives rise to a B-field in ten dimensions, which has a
component along the remaining $S^4$ isometry coordinate. Therefore
the background matrix (\ref{background}) is not symmetric anymore.
 On the other hand, reducing along the $AdS_4$ coordinate
$\phi_1$, one ends up with no B-field in ten dimensions and the
background matrix is still symmetric. As a result, the
transformation of the background matrices  under T-duality will
give results of different forms. In fact, each step of dimensional
reduction, T-duality and lifting, works differently for these two
options. In what follows, we analyze these steps separately for
each case and reach the (nontrivial but expected) result that both
choices yield the same deformed solution up to a sign in $\gamma$
terms.

Suppose that we reduce along one of the isometry directions of
$S^4$, say $\chi_2$. Then, we generate a B-field in 10 dimensions,
which has one leg along the isometry direction $\chi_4$:
$B_{\chi_1 \chi_4} \neq 0$ or $B_{\chi_3 \chi_4} \neq 0$. Thus, we
cannot apply the general formulas we derived before and have to
analyze the reduction, T-duality, lifting process again. After
reducing along $\chi_2$, the ten dimensional metric in the string
frame becomes: \be \frac{1}{\cos\chi_1} ds_{10 \ s}^2 =
-(1+r^2)dt^2 + \frac{dr^2}{1+r^2} + r^2d\Omega_5^2 + d\chi_1^2 +
\sin^2\chi_1[d\chi_3^2 + \sin^2\chi_3 d\chi_4^2] \, , \ee where we
used the fact that $A =0$ and $e^{4/3 \phi} = \cos^2\chi_1$.

Using (\ref{ansatzF}) and (\ref{ansatzF2}) the reduction of the
4-form field (\ref{f4dipole}) gives \be F_4  =  0, \ \ \ \ \ \ F_3
= dB = \cos\chi_1 \sin^2\chi_1 \sin\chi_3 d\chi_1  \wedge d\chi_3
\wedge d\chi_4 \, , \ee in ten dimensions. We choose our gauge
such that\footnote{This choice ensures that the B-field is
independent of the T-duality coordinate $\chi_4$, which is
essential.} \be \label{gaugeB} B = \frac{1}{6}\sin^3\chi_1
\sin\chi_3 d\chi_3 \wedge d\chi_4 + \frac{1}{2} \cos\chi_1
\sin^2\chi_1 \cos\chi_3 d\chi_1 \wedge d\chi_4. \ee Now our
background matrix (\ref{background}) is $6 \times 6$ and given as
\be \label{backdipole} E= \cos \chi_1 \left(\begin{array}{cccccc}
\sin^2\chi_1 \sin^2\chi_3 & 0 & 0 & 0 & -\frac{B_{\chi_1
\chi_4}}{\cos\chi_1} & -\frac{B_{\chi_3
\chi_4}}{ \cos\chi_1} \\
. & r^2 s_{\alpha}^2 & r^2 s_{\alpha}^2 c^2_{\theta} & r^2
s_{\alpha}^2 \cos2\theta & 0 & 0 \\
. & . & r^2(c_{\alpha}^2 + s_{\alpha}^2 c^2_{\theta}) &
r^2(-c_{\alpha}^2 + s_{\alpha}^2 c^2_{\theta}) & 0 & 0 \\
. & . & . & r^2 & 0 & 0 \\
\frac{B_{\chi_1 \chi_4}}{\cos\chi_1} & 0 & 0 & 0 & 1 & 0
\\
\frac{B_{\chi_3 \chi_4}}{\cos\chi_1} & 0 & 0 & 0 & 0 &
\sin^2\chi_1
\end{array}\right) \ee
in which the dotted entries are filled out by using the fact that
the first 4 columns and rows form a symmetric $4 \times 4$ matrix.
Here we have labeled our coordinates such that $x^1 = \chi_4, x^2
= \phi_1, x^3 = \phi_2, x^4 = \phi_3, x^5 = \chi_1, x^6 = \chi_3$.
Then we find: \bea \tilde{g}_{\mu \nu}& =& g_{\mu \nu}, \ \ \ \
\mu, \nu \in \{r, t,
\theta, \alpha\} \nonumber \\
\tilde{g}_{wi} &=& G g_{wi}, \ \ \ \ w=1,2, \ \ \ i = 1,2,3,4
\nonumber \\
\tilde{g}_{i5} &=& -\gamma G B_{15} g_{i2} \nonumber \\
\tilde{g}_{i6} &=& -\gamma G B_{16} g_{i2} \nonumber \\
\tilde{g}_{33} &=& G [g_{33} + \gamma^2 det E(4,5,6 \mid 4,5,6)]=
G[g_{33} +  \gamma^2\Delta r^2 \cos\chi_1 (\cos^2\alpha +
\frac{1}{4}\sin^2\alpha \sin^22\theta )]  \nonumber \\
\tilde{g}_{34} &=& G [g_{34} + \gamma^2 det E(3,5,6 \mid 4,5,6)] =
G[g_{34} + \gamma^2\Delta r^2 \cos\chi_1 (-\cos^2\alpha +
\frac{1}{2}\sin^2\alpha \sin^22\theta)] \nonumber \\
\tilde{g}_{44} &=& G [g_{44} + \gamma^2 det E(3,5,6 \mid 3,5,6)] =
G[g_{44} + \gamma^2\Delta r^2 \cos\chi_1 (\cos^2\alpha +
\sin^2\alpha \sin^22\theta)] \nonumber \\
\tilde{g}_{st} &=& g_{st} + G \gamma^2 B_{1s} B_{1t} g_{22} = G
g_{st} + G \gamma^2 (\Delta g_{st} + B_{1s} B_{1t} g_{22}) , \ \ \
s,t = 5,6 \eea and \bea \tilde{B}_{ij} &=& G \gamma (g_{i1}g_{2j}
- g_{i2}g_{1j}), \ \ \ i,j = 1,2,3,4 \nonumber \\
\tilde{B}_{1s} &=& G B_{1s}, \ \ \ s=5,6 \nonumber \\
\tilde{B}_{2s} &=& 0, \ \ \  \nonumber \\
\tilde{B}_{3s} &=& G \gamma^2 B_{1s}(g_{12}g_{23} - g_{22}g_{13})
\nonumber \\
\tilde{B}_{4s} &=& G \gamma^2 B_{1s}(g_{12}g_{24} - g_{22}g_{14})
\nonumber \\
\tilde{B}_{56} &=& 0, \eea where \be \label{bdipole} B_{15} =
-B_{\chi_1 \chi_4} = -\frac{1}{2} \sin^2\chi_1 \cos \chi_1
\cos\chi_3, \ \ \  B_{16} = - B_{\chi_3 \chi_4} =
-\frac{1}{6}\sin^3\chi_1 \sin\chi_3 \ee and \be G = (1 + \gamma^2
\Delta)^{-1} \ee with \be \label{deldipole} \Delta = g_{11}g_{22}
- g_{12}^2 = r^2 \sin^2\alpha
 \cos^2\chi_1 \sin^2\chi_1 \sin^2\chi_3. \ee

Looking at (\ref{f2f4}) we see that
$$\tilde{F}_2 = F_2 = dA = 0, \ \ \ \ \tilde{F}_4 = 0.$$ Then the
lifting of the metric to eleven dimensions with $\tilde{A} = 0 $
is given by: \bea  \label{type1metric} d\tilde{s}_{11}^2 & = &
e^{-2/3 \tilde{\phi}} d\tilde{s}_{10}^2 + e^{4/3 \tilde{\phi}}
d\chi_2^2
\nonumber \\
& = & G^{2/3} [ ds_{11}^2 + \gamma^2 e^{-2/3 \phi} \{det E(4,5,6
\mid 4,5,6) d\phi_2^2 + det E(3,5,6 \mid 4,5,6) d\phi_2 d\phi_3
\nonumber \\ & +&  det E(3,5,6 \mid 3,5,6)d\phi_3^2  +
\Delta((-1+r^2)dt^2 + \frac{dr^2}{1+r^2} + r^2 d\alpha^2 + r^2
\sin^2\alpha d\theta^2) \nonumber \\ &+& (\Delta g_{55} + B_{15}^2
g_{22}) d\chi_1^2 + (\Delta g_{66} + B_{16}^2 g_{22}) d\chi_3^2 +
(\Delta g_{56} + B_{15}B_{16} g_{22})
d\chi_1 d\chi_3 \} \nonumber \\
&-& \gamma e^{-2/3 \phi} \{(g_{12} d\chi_4 + g_{22} d\phi_1 +
g_{32} d\phi_2 + g_{42} d\phi_3)(B_{15} d\chi_1 + B_{16} d\chi_3)
\}], \eea where we have used $\tilde{g}_{\mu \nu} = g_{\mu \nu} =
G g_{\mu \nu} + (1-G)g_{\mu \nu} = G g_{\mu \nu} + \gamma^2 \Delta
G g_{\mu \nu}, \ \ \ \ \mu, \nu \in \{r, t, \theta, \alpha\}$. The
metric components can be read from (\ref{backdipole}).

Meanwhile, the only contribution to the deformed 4-form field in
eleven dimensions comes from the deformed B-field in ten
dimensions and is given by: \bea \label{ads7f2}
\tilde{\hat{F}}_4 &=& d(\tilde{B} \wedge d\chi_2)  \\
&=& d[G\sin^2\chi_1 (-\frac{1}{2}\cos \chi_3\cos \chi_1  d\chi_1
-\frac{1}{6} \sin \chi_1 \sin \chi_3 d \chi_3)\wedge  d\chi_2\wedge d\chi_4  \nonumber \\
&& + \gamma G  \Delta ( d\phi_1 + \cos^2 \theta  d \phi_2 + \cos 2
\theta  d\phi_3)]\wedge d\chi_2 \wedge d\chi_4  \nonumber \eea One
can check by explicit computation that if we interchange $x^1
\leftrightarrow x^2$, that is, if we set $x^1 = \phi_1, x^2 =
\chi_4, x^3 = \phi_2, x^4 = \phi_3, x^5 = \chi_1, x^6 = \chi_3$,
the eleven dimensional deformed metric and the deformed four form
field will be the same except for the fact that the $\gamma$
corrections will have an overall $-1$ factor. On the other hand,
if we start by reducing along $\chi_4$ and label the coordinates
of the background matrix such that $x^1= \phi_1, x^2=\chi_2$ we
again obtain the deformed metric (\ref{type1metric}) and the
deformed $\tilde{\hat{F}}_4$ (\ref{ads7f2}), whereas switching
$x^1$ with $x^2$ brings an overall -1 factor to the $\gamma$
corrections.

Now let us consider the case of dimensional reduction along the
isometry direction from $AdS_7$, i.e. $\phi_1$. This choice
generates  no B-field in 10 dimensions: \be \hat{F}_4 = F_4, \ \ \
\ B = 0. \ee After reduction along $\phi_1$ we obtain the ten
dimensional metric as \be \frac{1}{r\sin \alpha} ds_{10 \ s}^2 =
-(1+r^2)dt^2 + \frac{dr^2}{1+r^2} + r^2 d\bar{\Omega}_4 +
d\Omega_4, \ee where $\Omega_4$ is given in (\ref{sphere}) and
\bea d\bar{\Omega}_4 & = &  d\alpha^2 + \sin^2\alpha d\theta^2 +
(\cos^2\alpha + \frac{1}{4}\sin^2\alpha \sin^22\theta ) d\phi_2^2
\nonumber \\ & & + (\cos^2\alpha + \sin^2\alpha \sin^22\theta)
d\phi_3^2 + 2(-\cos^2\alpha + \frac{1}{2}\sin^2\alpha
\sin^22\theta) d\phi_2 d\phi_3. \eea Note that we have $e^{4/3
\phi} = r^2 \sin^2\alpha$ and $A = \cos^2\theta d\phi_2 +
\cos2\theta d\phi_3.$ Now we T-dualize along the remaining two
isometries: $\chi_2$ and $\chi_4$. Because these two coordinates
do not mix with any other direction in the metric, the background
matrix (\ref{background}) is only $2 \times 2$: \be E =
r\sin\alpha \left(\begin{array}{cc}
\cos^2\chi_1 & 0 \\
0 & \sin^2\chi_1 \sin^2\chi_3 \end{array}\right). \ee Acting on
this background matrix with the T-duality matrix (\ref{matrix})
with $d=2$ we obtain the deformed fields in ten dimensions: \be
\label{5000} \tilde{g}_{\chi_u \chi_v} = G g_{\chi_u \chi_v}, \ \
\ u,v = 2,4 \ \ \ \ \ \tilde{B}_{\chi_2 \chi_4} = \gamma G r^2
\sin^2\alpha \cos^2\chi_1 \sin^2\chi_1 \sin^2\chi_3, \ee where \be
G = (1 + \gamma^2 r^2 \sin^2\alpha \sin^2\chi_1 \cos^2\chi_1
\sin^2\chi_3 )^{-1}= (1 + \gamma^2 \Delta)^{-1}. \ee

We have a novelty in lifting this ten dimensional deformed metric
back to eleven dimensions. One can see from (\ref{f2f4}) that in
this case the two form $F_2$ has a nontrivial transformation and
lifting the 10 dimensional deformed metric to eleven dimensions
one has to use the deformed one-form $\tilde{A}$. \bea
\label{ads7f} \tilde{F}_2 & = & F_2 - \gamma i_{ \chi_2} i_{
\chi_4} F_4 \nonumber \\& = &
\sin2\theta(d\phi_2 \wedge d\theta + 2 d\phi_3 \wedge d\theta) + \gamma \cos\chi_1 \sin^2\chi_1 \sin\chi_3 d\chi_1 \wedge d\chi_3, \nonumber \\
\tilde{F}_4 & = & F_4 + \tilde{F}_2 \wedge \tilde{B}. \eea In a
suitable gauge consistent with the choice we made in
(\ref{gaugeB}), the deformed one-form $\tilde{A}$ to be used in
lifting the ten dimensional deformed metric is \bea
\label{adipole} \tilde{A} &=& \cos^2\theta d\phi_2 + \cos2\theta
d\phi_3 + \gamma (\frac{1}{6}\sin^3\chi_1 \sin\chi_3 d\chi_3 +
\frac{1}{2} \cos\chi_1 \sin^2\chi_1 \cos\chi_3 d\chi_1) \nonumber \\
& \equiv & A + \gamma \cal{A}. \eea Then we get \bea
d\tilde{s}_{11}^2 & = & e^{-2/3 \tilde{\phi}} d\tilde{s}_{10}^2 +
e^{4/3 \tilde{\phi}} (d\phi_1 + \tilde{A})^2
\nonumber \\
& = & G^{2/3}[(1+\gamma^2 \Delta) e^{-2/3 \phi} d\tilde{s}_{10}^2
+ e^{4/3 \phi} (d\phi_1 + A + \gamma {\cal{A}})^2] \eea where we
have used $G^{-1} = 1 + \gamma^2 \Delta$.  Also using \be
d\tilde{s}^2_{10} =  ds^2_{10} - \gamma^2 \Delta (g_{\chi_2
\chi_2}d\chi_2^2 + g_{\chi_4 \chi_4}d\chi_4^2) \ee we find \bea
\label{type2metric} d\tilde{s}_{11}^2 & = & G^{2/3}[ds^2_{11} +
e^{-2/3 \phi} \gamma^2 \Delta (ds^2_{10} - g_{\chi_2
\chi_2}d\chi_2^2 - g_{\chi_4
\chi_4}d\chi_4^2) \nonumber \\
&&+ e^{4/3 \phi} \gamma^2 {\cal{A}}^2 + e^{4/3 \phi} \gamma
(d\phi_1 + A){\cal{A}}] \eea Reading $A$ and $\cal{A}$ from
(\ref{adipole}) and using $e^{4/3 \phi} = r^2 \sin^2\alpha$ one
finds that the metric in (\ref{type2metric}) is exactly the same
as the metric in (\ref{type1metric}). Here too, changing the order
of $\chi_2$ and $\chi_4$ brings an overall -1 sign to $\cal{A}$
(as the two contractions $i_{ \chi_2}$ and $i_{ \chi_4}$
anticommute) and hence an overall -1 sign to the $\gamma$
correction to the eleven dimensional metric in
(\ref{type2metric}). The deformed 4-form field  in 11-dimensions
can be found from (\ref{ansatzF}) \bea \label{5001}
\tilde{\hat{F}}_4 &=& \tilde{F}_4 + d\tilde{B} \wedge (d\phi_1 + \tilde{A})  \\
 &=& \hat{F}_4 + \gamma d[G\Delta ( d\phi_1 +
\cos^2 \theta   d\phi_2 +
\cos 2 \theta  d\phi_3)] \wedge d\chi_2 \wedge d\chi_4  \nonumber \\
&& + \gamma^2 d[G\Delta(\frac{1}{6} \sin^3\chi_1\sin \chi_3
d\chi_3 + \frac{1}{2} \cos \chi_1\sin^2 \chi_1  d\chi_1)] \wedge
d\chi_2 \wedge d\chi_4 \nonumber \eea where we use (\ref{5000}),
(\ref{ads7f}) and (\ref{adipole}). It can be shown that
(\ref{5001}) is equivalent to (\ref{ads7f2}) by using the identity
$ G= 1- \gamma^2 \Delta G$. Again interchanging $x^1
\leftrightarrow x^2$ brings an overall -1 factor to the $\gamma$
corrections in (\ref{5001}).

Therefore, it is still true that once we fix $T^3$ (which, in this
case is the $\{\chi_2, \chi_4, \phi_1\}$ torus), it does not make
a difference as to how we choose the reduction and T-duality
directions, up to a sign in the $\gamma$ corrections to the metric
and the 4-form field as was observed in \cite{berman}. We see from
the above discussions that the sign is the same as the sign of
$\epsilon^{zx^1x^2}$, where $\epsilon^{\chi_4\chi_2 \phi_1}=1$.
Note that this sign issue in the metric did not arise in the
previous subsections, as we had symmetric background matrices and
hence no $\gamma$ corrections in the eleven dimensional deformed
metric.

\section{Multiparameter Deformations}

In the previous section, we studied 1-parameter deformations of
M-theory backgrounds with five commuting isometries. Our method
involved fixing a three dimensional subtorus  of the 5-torus
associated with these isometries. Then we  used one of the
directions of this 3-torus for the reduction and the remaining two
for  T-dualization. In this section, we will generalize our
discussion to multiparameter deformations. The most general
deformation would have ten parameters, which can be obtained by
performing our method ten times subsequently, by using the
$\left(\begin{array}{c} 5 \\ 3
\end{array}\right) = 10$ possible 3-tori embedded in the 5-torus. A less complicated, 6-parameter deformation can
be obtained by fixing the reduction direction, say $z$, and then
applying  the T-duality transformation via the matrix
(\ref{matrix}), $\left(\begin{array}{c} 4 \\ 2
\end{array}\right) = 6$ times. As we have seen, once the 3-torus is decided, it does not make a
difference as to which direction is chosen for the reduction.
Therefore, the 6-parameter deformation obtained  this way would be
equivalent to a deformation by using the 6 possible 3-tori with
coordinates $\{z, x^i, x^j\}$, where $x^i, x^j$ are chosen from
the 4 remaining isometries of $T^5$. The six consecutive T-duality
transformations in ten dimensions can still be obtained by the
action of a single  $O(4,4)$ matrix $T$ defined as \cite{aybike}:
\be \label{product} T = T_1 . T_2 . T_3 . T_4 . T_5 . T_6 =
\left(\begin{array}{cc}
1 & 0 \\
\Gamma & 1
\end{array}\right)
\ee where each $T_r$ is an $O(4,4)$ matrix of the form \be T_r =
\left(\begin{array}{cc}
1 & 0 \\
\Gamma_r & 1
\end{array}\right), \ \ \ \ r=1,\cdots,6 \ee with
\be \Gamma_1 = \left(\begin{array}{cccc} 0 & -\gamma_3 & 0 & 0\\
\gamma_3 & 0 & 0 & 0\\
0 & 0 & 0 & 0 \\
0 & 0 & 0 & 0 \end{array}\right), \ \ \ \ \Gamma_2 = \left(\begin{array}{cccc} 0 & 0 & \gamma_2 & 0 \\
0 & 0 & 0 & 0 \\
-\gamma_2 & 0 & 0 & 0 \\ 0 & 0 & 0 & 0 \end{array}\right), \ \ \ \ \Gamma_3 = \left(\begin{array}{cccc} 0 & 0 & 0 & 0 \\
0 & 0 & -\gamma_1 & 0 \\
0 & \gamma_1 & 0 & 0 \\ 0 & 0 & 0 & 0 \end{array}\right). \ee

\be \Gamma_4 = \left(\begin{array}{cccc} 0 & 0 & 0   & -\gamma_4 \\
0 & 0 & 0 & 0\\
0 & 0 & 0 & 0 \\
\gamma_4 & 0 & 0 & 0 \end{array}\right), \ \ \ \ \Gamma_5 = \left(\begin{array}{cccc} 0 & 0 & 0 & 0 \\
0 & 0 & 0 & \gamma_5 \\
0 & 0 & 0 & 0 \\ 0 & -\gamma_5 & 0 & 0 \end{array}\right), \ \ \ \ \Gamma_6 = \left(\begin{array}{cccc} 0 & 0 & 0 & 0 \\
0 & 0 & 0 & 0 \\
0 & 0 & 0 & -\gamma_6 \\ 0 & 0 & \gamma_6 & 0 \end{array}\right).
\ee From (\ref{product}) it is easy to see that $\Gamma = \Gamma_1
+ \Gamma_2 + \Gamma_3 + \Gamma_4 + \Gamma_5 + \Gamma_6 $. Setting
$\gamma_1 = \gamma_2 = \gamma_4 = \gamma_5 = \gamma_6 =0$, $T$
given in (\ref{product}) becomes equal to the T-duality matrix in
(\ref{matrix}) that we used for the one-parameter deformation with
$\gamma = \gamma_3$. Instead of giving the results for a
6-parameter deformation, we will make a further simplification and
set $\gamma_4 = \gamma_5 = \gamma_6 =0$, that is, we will not
involve the isometry corresponding to the shift of the coordinate
$x^4$ in the deformation process. This is a sensible choice to
make, as one of the isometries in the original eleven dimensional
background will always be associated with the R-symmetry on the
field theory side \footnote{In all but one of the examples below
we will omit the isometry that corresponds to the R-symmetry, so
that the supersymmetry is not lost. However, in 2 or 3-parameter
$\beta$ deformations cases we will be forced to involve this
isometry. Then, the unused isometry will be the one coming from
the $AdS$ part of the geometry.}. Let us remind that, strictly
speaking $x^4$ does not have to be a $U(1)$ direction, since we
are not using it for deformation process. With this the matrix
$\Gamma$ becomes \be \label{6pgamma} \Gamma =
\Gamma_1 + \Gamma_2 + \Gamma_3 = \left( \begin{array}{cccc} 0 & -\gamma_3 & \gamma_2 & 0\\
\gamma_3 & 0 & -\gamma_1 & 0 \\
-\gamma_2 & \gamma_1 & 0 & 0 \\
0 & 0 & 0 & 0
\end{array}\right). \ee
This choice corresponds to performing 3 successive $O(2,2)$
T-duality transformations in $D=10$ using the isometry directions
$\{x^1,x^2\}, \{x^1,x^3\}$ and $\{x^2,x^3\}$ with parameters
$\gamma_3, \gamma_2, \gamma_1$ respectively. Each $O(2,2)$ leads
to a LM deformation.

We again assume that $\hat{F}_4$ has at most one leg along the
torus directions which are decoupled from the rest of the
geometry. With these assumptions the background matrix $E$ in
$D=10$ (\ref{background}) becomes symmetric. Now we reduce along
one of the coordinates which we have named $x^5 = z$ and deform
the resulting ten dimensional metric in the string frame using
(\ref{product}) on $E$. From the transformation rules
(\ref{onemli}) and (\ref{dilaton}) we find: \bea
\label{3pdeformed} \tilde{g}_{ab} & = & G[g_{ab} + \gamma_a
\gamma_b \ detg(4 \mid 4)], \ \ \ a,b =
1,2,3 \nonumber \\
\tilde{g}_{a4} & = & G[g_{a4} + \sum_{b=1,2,3} (-1)^{b+1} \
\gamma_a \gamma_b \ detg(b \mid 4)], \ \ \ a=1,2,3
\nonumber \\
\tilde{g}_{44} & = & G[g_{44} + \sum_{a,b = 1,2,3}  (-1)^{a+b} \
\gamma_a
\gamma_b \ detg(a \mid b)] \nonumber \\
\tilde{B}_{ij}& = & G \epsilon^{ij} \sum_{k,l=1,2,3,4;\ k,l \neq
i,j;\ k\neq l}\ \sum_{a=1,2,3} (-1)^{a+1} \gamma_a \ detg(k,l \mid
a,4) \nonumber
\\ e^{2 \tilde{\phi}} & = & G e^{2\phi}, \eea where $G =
det(\Gamma E + 1)^{-1}$ is \be \label{3pG} G = [1 +
\sum_{a,b=1,2,3} (-1)^{a+b} \ \gamma_a \gamma_b \ detg(4,a \mid
b,4) ]^{-1}. \ee  In order to lift this deformed ten dimensional
metric back to eleven dimensions we go through the same steps as
in the previous section and derive \bea \label{3pesassonuc}
d\tilde{s}_{11}^2 &=& G^{-1/3}\hat{g}_{\mu \nu} dx^{\mu} dx^{ \nu}
+ G^{2/3} \hat{g}_{mn} dx^m dx^n  \\ && + G^{2/3}
\sum_{a,b=1,2,3}[\gamma_a \gamma_b \ det\hat{g}(4 \mid 4) dx^{a}
dx^{b} + (-1)^{b+1} \ \gamma_a \gamma_b \ det\hat{g}(b \mid 4)
dx^{a} dx^{4}  \nonumber
\\ && +  (-1)^{a+b} \ \gamma_a \gamma_b \
det\hat{g}(a \mid b) dx^{4} dx^{4})], \nonumber \eea with \be
\label{3phatG} G^{-1} = 1 + \sum_{a,b=1,2,3} (-1)^{a+b} \ \gamma_a
\gamma_b \ det\hat{g}(4,a \mid b,4). \ee Here $\hat{g}$ is the $5
\times 5$ torus matrix with entries $\hat{g}_{mn}$.

On the other hand, the Ramond-Ramond fields in the deformed ten
dimensional geometry can be found via the action on (\ref{RR2}) of
the operator ${\bf T}$  (\ref{RR7}) which takes the form \be {\bf
T} = 1 - \frac{1}{2} \epsilon^{abc} \ \gamma_c \ i_{a} i_{b} \ee
for $\Gamma$ in (\ref{6pgamma}). Here $\epsilon^{123} = 1$.
Finding the deformed 4-form field and lifting it to eleven
dimensions with the B-field in the deformed geometry we find \bea
\label{3pdeformedF11asil} \tilde{\hat{F}}_4 &=& \hat{F}_4 -
\frac{1}{2} \epsilon^{abc} \ \gamma_c \ i_{a} i_{b} i_{z}
\star_{11}
 \hat{F}_4  \nonumber \\&& + \frac{(-1)^{a+1} \gamma_a}{2} d\left( G \sum_{q,r \neq m,n,p ; \ q \neq r} det\hat{g}(q,r \mid a,4)
 \frac{\epsilon^{mnp} dx_m \wedge dx_n \wedge dx_p}{3!}\right) \, ,
 \eea
where $\epsilon^{mnp}=1$ for $m<n<p$ with $m,n,p=\{1,...,5\}$ and
$a,b,c=\{1,2,3\}$. Note that in deriving (\ref{3pesassonuc}) and
(\ref{3pdeformedF11asil}) we used $\{x^1,x^2,x^5\}$,
$\{x^1,x^3,x^5\}$ and $\{x^2,x^3,x^5\}$ tori with deformation
parameters $\gamma_3, \gamma_2$ and $\gamma_1$ respectively.

Let us note that when $\gamma_2=\gamma_1=0$ then formulas
(\ref{3pesassonuc}), (\ref{3pdeformedF11asil}) and (\ref{3pG})
reduce to our single parameter results (\ref{esassonuc}),
(\ref{deformedF11asil}) and (\ref{G}) with $\gamma_3=\gamma$. When
only one of the $\gamma_a=0$ we have a 2-parameter deformation.
Finally we would like to point out that our formulas are still
true if the $x^4$ coordinate is not a $U(1)$ direction but just a
coordinate that mixes with $\{x^1,x^2,x^3,x^5\}$. If in the given
background there is no such mixing or if there are only 4
decoupled $U(1)$ directions to begin with, it is straightforward
to adopt our main formulas (\ref{3pesassonuc}),
(\ref{3pdeformedF11asil}) and (\ref{3pG}). This will just be a
special case of the above where the 4'th $U(1)$ direction does not
couple with others. Let $\hat{g}$ be the $4\times 4$ torus matrix
of $\{x^1,x^2,x^3,x^5\}$. Then, \bea d\tilde{s}_{11}^2 &=&
G^{-1/3}\hat{g}_{\mu \nu} dx^{\mu} dx^{ \nu} + G^{2/3}
\hat{g}_{mn} dx^m dx^n  + G^{2/3} \sum_{a,b=1,2,3} \gamma_a
\gamma_b \ det\hat{g} \, dx^{a} dx^{b} \nonumber\\
\tilde{\hat{F}}_4 &=& \hat{F}_4 -
 \frac{\epsilon^{abc}}{2} \ \gamma_c \ i_{a} i_{b} i_{z}
\star_{11} \hat{F}_4 + (-1)^{a+1} \gamma_a d\, [G \sum_{q \neq
m,n,p } det\hat{g}(q \mid a)
\frac{\epsilon^{mnp} dx_m \wedge dx_n \wedge dx_p}{3!}] \nonumber \\
\label{4U12} G^{-1} &=& 1 + \sum_{a,b=1,2,3} (-1)^{a+b} \ \gamma_a
\gamma_b \ det\hat{g}(a \mid b) , \eea where $m,n,p=\{1,2,3,5\}$.

\subsection{Example: $AdS_4 \times ({\rm Sasaki-Einstein})_7$ ({\rm with base} $S^2 \times S^2$)}

In this section we will apply the method we described above to
obtain 3-parameter deformations of the background
$$AdS_4 \times X_7$$ where $X_7$ is the seven dimensional
Sasaki-Einstein space found recently in \cite{sasaki2}  with base
$S^2 \times S^2$. The metric and 4-form field are \be ds_{11}^2 =
ds_{AdS_4}^2 + ds_{X_7}^2 \, , \hspace{1cm} \hat{F}_4= 6 \,
vol(AdS_4) \, ,  \label{fsasaki}\ee where $AdS_4$ metric is given
in (\ref{adsmetric}) and
 \begin{eqnarray} \label{s2base}
ds_{X_7}^2 &=& U^{-1}d\rho^2 + \frac{\rho^2}{2}(d\theta_1^2 +
\sin{\theta_1}^2d\phi_1^2 + d\theta_2^2 +
\sin{\theta_2}^2d\phi_2^2) \nonumber \\
&&+ q(d\psi+j_1)^2 + w[d\alpha + f(d\psi+j_1)]^2 \end{eqnarray}
Here the radius of the Sasaki-Einstein manifold is taken to be 1.
In this case the radius of the $AdS_4$ is 1/2 and in writing the
above metric we scaled time and radial coordinates with factor 2.
The functions $U, \o, f$ and $q$ are functions of $\r$ that are
given in \cite{sasaki2} and  $j_1 = -\cos{\theta_1} d\phi_1 -
\cos{\theta_2}d\phi_2$.

In addition to the $AdS_4$ Killing vector $\partial_{\chi_2}$
there are four more commuting isometries of $X_7$ with Killing
vectors $\partial_{\phi_1},
\partial_{\phi_2},
\partial_{\alpha},\partial_{\psi}
$ . The last one corresponds to the R-symmetry $U(1)$ on the field
theory side.

\subsubsection{$\beta$ Deformations}

We label the coordinates as $x^1 = \phi_1$, $x^2 = \phi_2, x^3 =
\psi$ and $x^4 = \chi_2$ and choose $\alpha$ as the reduction
direction, i.e.,  $z = x^5 = \alpha$. Setting $x^4 = \chi_2$
guarantees that the deformation involves no dipole deformation.
These correspond to the choice of three 3-tori with coordinates
$\{\phi_1,\phi_2,\alpha \}, \{\phi_1,\psi, \alpha\},
\{\phi_2,\psi, \alpha\}$, whose deformation parameters are
$\gamma_3, \gamma_2$ and $\gamma_1$, respectively. The  5-torus
matrix is ($s_i \equiv \sin{\theta_i}, \, c_i \equiv
\cos{\theta_i}$, \,  $i=1,2$)
\be \label{bgs2s2} \hat{g} =
\left(\begin{array}{ccccc} \frac{\r^2}{2}s_1^2+ (wf^2 + q) c_1^2 &
(wf^2 + q) c_1 c_2
& -(wf^2 + q) c_1 & 0 & -wfc_1 \\
. & \frac{\r^2}{2}s_2^2+ (wf^2 + q) c_2^2 & \ -(wf^2 + q) c_2 & 0 & -wfc_2 \\
. & . & \ (wf^2 + q) & 0 & wf \\
. & . & . & r^2 \sin^2\chi_1 & 0 \\
. & . & . & . & w
 \end{array}\right).
\ee
Missing entries are filled using the fact that $\hat{g}$ is a
symmetric matrix. Evaluating the relevant non-zero subdeterminants are given in (\ref{subdeterminant1}).
Then using (\ref{3pesassonuc}) the deformed metric becomes
\bea \label{deformed3psasaki}  ds_{11}^2 &=&
G^{-1/3}\{ds_{AdS_4}^2  + U^{-1}d\rho^2 +
\frac{\rho^2}{2}(d\theta_1^2  + d\theta_2^2 )\}  \\
&+& G^{2/3}\{\frac{\rho^2}{2}(s_1^2d\phi_1^2 + s_2^2d\phi_2^2)+
q(d\psi+j_1)^2 + w[d\alpha + f(d\psi+j_1)]^2 \} \nonumber \\
&+& G^{2/3} \frac{\o q \r^4 s_{1}^2s_{2}^2}{4} \{ \g_3 ^2 d\psi^2
 + (\g_1 d\phi_1 + \g_2 d\phi_2)^2
+ 2 \g_1 \g_3  d\phi_1 d\psi  + 2 \g_2 \g_3  d\phi_2 d\psi \}.
\nonumber \eea where \bea G^{-1} &=&1 + \frac{\gamma_3^2 \r^2
w}{4}[2q(s_1^2 c_2^2 + c_1^2 s_2^2) + \r^2  s_1^2 s_2^2] + \g_2^2
\frac{\r^2 \o q s_1^2}{2}+
\g_1^2 \frac{\r^2 \o q s_2^2}{2} \nonumber \\
&+& \g_1 \g_3   \r^2 \o q c_1 s_2^2 - \g_2 \g_3 \r^2 \o q c_2
s_1^2. \eea Note that in deriving (\ref{deformed3psasaki}) we used
(\ref{3pesassonuc}) in which the $d\chi_2^2$ terms combine to give
the $d\chi_2^2$ term in the $AdS_4$ metric (\ref{adsmetric})  by
using the fact that $G^{-1/3} = G^{2/3} G^{-1}$. On the other hand
from (\ref{fsasaki}) we get \be \star_{11}\hat{F}_4 = 6 vol(X_7) =
\frac{3\r^4 s_1 s_2 (q \o)^{1/2}}{2U^{1/2}} d\r \wedge d\theta_1
\wedge d\theta_2 \wedge d\psi \wedge d\phi_1 \wedge d\phi_2 \wedge
d\alpha \, , \ee and using (\ref{3pdeformedF11asil}) we find the
deformed 4-form field $\tilde{\hat{F}}_4$ as \bea \label{sasakif}
\tilde{\hat{F}}_4 & = & \hat{F}_4 - (\frac{9q \o}{4 U})^{1/2}\r^4
s_1 s_2 d\r \wedge d\theta_1
\wedge d\theta_2 \wedge [ \g_1 d\phi_1 + \g_2 d\phi_2    + \g_3  d\psi ]  \nonumber \\
&+& \g_3d(G\, [\frac{\r^2 w}{4}(2q(s_1^2 c_2^2 + c_1^2 s_2^2) +
\r^2 s_1^2 s_2^2) d\phi_1 \wedge d\phi_2 \wedge d\alpha +
\frac{1}{2}\r^2 \o q c_1 s_2^2 d\phi_2 \wedge d\psi \wedge d\alpha
\nonumber \\ &-& \frac{1}{2}\r^2 \o q c_2 s_1^2 d\phi_1 \wedge
d\psi \wedge d\alpha +\frac{1}{4} \r^4 \o f s_1^2 s_2^2 d\phi_1
\wedge d\phi_2 \wedge d\psi]) \nonumber \\
&-& \g_2 d(G\, [ -\frac{1}{2}\r^2 \o q c_2 s_1^2 d\phi_1 \wedge
d\phi_2 \wedge d\alpha + \frac{1}{2}\r^2 \o q s_1^2 d\phi_2 \wedge
d\psi
\wedge d\alpha]) \nonumber \\
&+& \g_1 d(G\, [\frac{1}{2}\r^2 \o q c_1 s_2^2 d\phi_1 \wedge
d\phi_2 \wedge d\alpha + \frac{1}{2}\r^2 \o q s_2^2  d\phi_2
\wedge d\psi \wedge d\alpha]). \eea Here we used the orientation
given in \cite{jerome} for the volume forms which are fixed by
Killing spinors of $X_7$. Setting $\gamma_1 = \gamma_2 = 0$  and
$\gamma_3=\gamma$ in (\ref{deformed3psasaki}) and (\ref{sasakif})
we find the result of \cite{jerome} albeit presented in a
different way. Note that this is the only $\beta$ deformation for
which the R-symmetry $\partial_{\psi}$ is left untouched.

\subsubsection{Dipole Deformations}
In this case one of the coordinates of the 3-tori used for the
deformation should always be chosen as the $AdS_4$ isometry
direction $\chi_2$. We ensure that by choosing $x^5 = z = \chi_2$.
The omitted isometry corresponds to the shift of the R-symmetry
direction $\psi$, so we set $x^4 = \psi$. Then our 5-torus matrix
is obtained from (\ref{bgs2s2}) by interchanging the columns and
rows in the following order: $4 \leftrightarrow 5, 3
\leftrightarrow 4$ . \be \label{bgs2s2dipole} \hat{g} =
\left(\begin{array}{ccccc} \frac{\r^2}{2}s_1^2+ (wf^2 + q) c_1^2 &
(wf^2 + q) c_1 c_2 & -wfc_1
& -(wf^2 + q) c_1  & 0 \\
. & \frac{\r^2}{2}s_2^2+ (wf^2 + q) c_2^2 & -wfc_2 & \ -(wf^2 + q) c_2 & 0  \\
. & . & w & wf & 0 \\
. & . & . & (wf^2 + q) & 0 \\
. & . & . & . & r^2 \sin^2\chi_1
 \end{array}\right). \ee
Note that we have set $x^1 = \phi_1, x^2 = \phi_2, x^3 = \alpha$.
Therefore the 3-parameter deformation we will present here is the
one obtained by using the 3-tori $\{\chi_2, \phi_1, \phi_2\}$,
$\{\chi_2, \phi_1, \alpha\}$ and $\{\chi_2, \phi_2, \alpha\}$. The
relevant non-zero subdeterminants are given in (\ref{subdeterminant}).
So, we see from
(\ref{3pesassonuc}) that the deformed metric becomes  \bea
\label{deformed3psasakidipole} ds_{11}^2 &=&
G^{-1/3}\{ds_{AdS_4}^2 - r^2 \sin^2\chi_1 d\chi_2^2 +
U^{-1}d\rho^2 +
\frac{\rho^2}{2}(d\theta_1^2  + d\theta_2^2 )\}  \\
&+& G^{2/3}\{\frac{\rho^2}{2}(s_1^2d\phi_1^2 + s_2^2d\phi_2^2)+
q(d\psi+j_1)^2 + w[d\alpha + f(d\psi+j_1)]^2 + r^2 \sin^2\chi_1 d\chi_2^2\} \nonumber \\
&+& G^{2/3} r^2 \sin^2\chi_1 \{ \g_3 ^2 [\frac{\r^2
\o}{4}(2q(s_1^2 c_2^2 + c_1^2 s_2^2) + \r^2 s_1^2 s_2^2) d\alpha^2
+ \frac{1}{4} \r^4 \o f
s_1^2 s_2^2 d\alpha d\psi \nonumber \\
&+& \frac{1}{4}\r^4   (q + \o f^2) s_1^2 s_2^2 d\psi^2] \nonumber
\\ &+& \g_2^2 [\frac{\r^2 \o}{4}(2q(s_1^2 c_2^2 + c_1^2 s_2^2) +
\r^2 s_1^2 s_2^2) d\phi_2^2 - \frac{1}{2}\r^2  \omega q  c_2 s_1^2
d\phi_2 d\psi + \frac{1}{2}\r^2  \omega q  s_1^2 d\psi^2]
\nonumber \\
&+&  \g_1^2  [\frac{\r^2 \o}{4}(2q(s_1^2 c_2^2 + c_1^2 s_2^2) +
\r^2 s_1^2 s_2^2)d\phi_1^2 -\frac{1}{2}\r^2  \omega q  c_1 s_2^2
d\phi_1 d\psi + \frac{1}{2}\r^2  \omega q  s_2^2 d\psi^2] \nonumber \\
&+& \g_1 \g_2   [\frac{\r^2 \o}{2}(2q(s_1^2 c_2^2
+ c_1^2 s_2^2) + \r^2 s_1^2 s_2^2) d\phi_1 d\phi_2 -\frac{1}{2}\r^2  \omega q  c_1 s_2^2 d\phi_2 d\psi - \frac{1}{2}\r^2  \omega q  c_2 s_1^2 d\phi_1 d\psi] \nonumber \\
&+& \g_1 \g_3 [\frac{\r^2 \o}{2}(2q(s_1^2 c_2^2
+ c_1^2 s_2^2) + \r^2 s_1^2 s_2^2)d\phi_1 d\alpha + \frac{1}{4} \r^4 \o f s_1^2 s_2^2 d\phi_1 d\psi -\frac{1}{2}\r^2  \omega q  c_1 s_2^2 d\alpha d\psi] \nonumber \\
&+& \g_2 \g_3  [\frac{\r^2 \o}{2}(2q(s_1^2 c_2^2 + c_1^2 s_2^2) +
\r^2 s_1^2 s_2^2)d\phi_2 d\alpha + \frac{1}{4} \r^4 \o f s_1^2
s_2^2 d\phi_2 d\psi -\frac{1}{2}\r^2  \omega q  c_2 s_1^2 d\alpha
d\psi]\}, \nonumber \eea where \bea G^{-1} &= &1 + r^2
\sin^2\chi_1[ \gamma_3^2 \frac{\r^2 }{4}[2(q + \o f^2)(s_1^2 c_2^2
+ c_1^2 s_2^2) + \r^2 s_1^2 s_2^2] + \g_2^2 \frac{\o}{2}  (2q
c_1^2 + \r^2 s_1^2) \nonumber \\ &+& \g_1^2 \frac{1}{2}  \o (2q
c_2^2 + \r^2 s_2^2) -2\g_1 \g_2 q \o c_1 c_2 +\g_1 \g_3 \r^2 \o f
c_1 s_2^2 + \g_2 \g_3 \r^2 \o f c_2 s_1^2].  \eea After the
deformation the 4-form field can be found using
(\ref{3pdeformedF11asil}) as: \bea \tilde{\hat{F}}_4&=&
\label{3dipolef}
\hat{F}_4 + d\phi_1 \wedge d\psi \wedge d\chi_2 \wedge d(G\, [\gamma_3 det\hat{g}(2,3 \mid 3,4) - \gamma_2 det\hat{g}(2,3 \mid 2,4)]) \nonumber \\
&+& d\phi_2 \wedge d\alpha \wedge d\chi_2 \wedge d(G\, [\gamma_1
det\hat{g}(1,4 \mid 1,4)
- \gamma_2 det\hat{g}(1,4 \mid 2,4) + \gamma_3 det\hat{g}(1,4 \mid 3,4)]) \nonumber \\
&+& d\phi_1 \wedge d\phi_2 \wedge d\chi_2 \wedge d(G \, [\gamma_1
det\hat{g}(1,4 \mid 3,4)
- \gamma_2 det\hat{g}(2,4 \mid 3,4) + \gamma_3 det\hat{g}(3,4 \mid 3,4) ]) \nonumber \\
&+& d\phi_1 \wedge d\alpha \wedge d\chi_2 \wedge d(G\, [\gamma_1
det\hat{g}(1,4 \mid 2,4)
- \gamma_2 det\hat{g}(2,4 \mid 2,4) + \gamma_3 det\hat{g}(2,4 \mid 3,4)]) \nonumber \\
&+& d \alpha \wedge d\psi \wedge d \chi_2 \wedge d(G \, [ \gamma_1
det\hat{g}(1,2 \mid 1,4)
-\gamma_2 det\hat{g}(1,2 \mid 2,4)]) \nonumber \\
&+& d\phi_2 \wedge d\psi \wedge d\chi_2 \wedge d(G\, [ \gamma_1
det\hat{g}(1,3 \mid 1,4)
 + \gamma_3 det\hat{g}(1,3 \mid 3,4)]) \, .
\eea where necessary determinants are given in
(\ref{subdeterminant}). Note that there is no contribution to the
above from the Hodge dual $*_{11}\hat{F}_4$ since it has no leg in
the reduction direction $\chi_2$.

\subsubsection{Mixed Deformations}
Here we will consider a multiparameter deformation where a
combination of dipole and $\beta$ deformations are applied. Our
choice for the reduction coordinate is $z = x^5 = \alpha$. We set
$x^1 = \phi_1, x^2 = \phi_2, x^3 = \chi_2, x^4 = \psi, x^5 =
\alpha$. Therefore the deformation is obtained by using the 3-tori
$\{\alpha, \phi_1, \phi_2\}$, $\{\alpha, \phi_1, \chi_2\}$ and
$\{\alpha, \phi_2, \chi_2\}$ and it is a mixed deformation
involving two dipole and one $\beta$ deformation. The R-symmetry
direction $\psi$ is not used\footnote{Note that there are two more
possibilities. Had we chosen $x^1 = \alpha, x^2 = \phi_2, x^3 =
\chi_2, x^4 = \psi, x^5 = \phi_1$ we would have a deformation
obtained by using the 3-tori $\{\phi_1, \alpha, \phi_2\}$,
$\{\phi_1, \alpha, \chi_2\}$ and $\{\phi_1, \phi_2, \chi_2\}$.
Similarly choosing $x^1 = \phi_1, x^2 = \alpha, x^3 = \chi_2, x^4
= \psi, x^5 = \phi_2$ would give a deformation obtained by using
the 3-tori $\{\phi_2, \phi_1, \alpha\}$, $\{\phi_2, \phi_1,
\chi_2\}$ and $\{\phi_2, \alpha, \chi_2\}$.}. The 5-torus matrix
is obtained from (\ref{bgs2s2dipole}) by interchanging the 3rd and
5th columns and rows: \be \label{bgs2s2mixed} \hat{g} =
\left(\begin{array}{ccccc} \frac{\r^2}{2}s_1^2+ (wf^2 + q) c_1^2 &
(wf^2 + q) c_1 c_2 & 0
& -(wf^2 + q) c_1  & -wfc_1 \\
. & \frac{\r^2}{2}s_2^2+ (wf^2 + q) c_2^2 & 0 & \ -(wf^2 + q) c_2 & -wfc_2  \\
. & . & r^2 \sin^2\chi_1 & 0 & 0 \\
. & . & . & (wf^2 + q) & wf \\
. & . & . & . & w
 \end{array}\right). \ee
The relevant subdeterminants which are non-zero are given in(\ref{subdeterminant2}).
So, we see from
(\ref{3pesassonuc}) that the deformed metric becomes \bea
\label{mixedmetric} ds_{11}^2 &=& G^{-1/3}\{ds_{AdS_4}^2 - r^2
\sin^2\chi_1 d\chi_2^2 + U^{-1}d\rho^2 +
\frac{\rho^2}{2}(d\theta_1^2  + d\theta_2^2 )\}  \\
&+& G^{2/3}\{\frac{\rho^2}{2}(s_1^2d\phi_1^2 + s_2^2d\phi_2^2)+
q(d\psi+j_1)^2 + w[d\alpha + f(d\psi+j_1)]^2 + r^2 \sin^2\chi_1 d\chi_2^2 \nonumber \\
&+& \frac{\g_3^2}{4}\r^4  q \omega   s_1^2 s_2^2 d\psi^2\} \nonumber \\
&+& G^{2/3} r^2 \sin^2\chi_1 \{ \g_3^2[\frac{\r^2 \o}{4} (2q(s_1^2
c_2^2 + c_1^2 s_2^2) + \r^2 s_1^2 s_2^2) d\chi_2^2]
\nonumber \\
&+&  \g_1^2  [\frac{\r^2 \o}{4}(2q(s_1^2 c_2^2 + c_1^2 s_2^2) +
\r^2 s_1^2 s_2^2)d\phi_1^2 -\frac{1}{2}\r^2  \omega q  c_1 s_2^2
d\phi_1 d\psi + \frac{1}{2}\r^2  \omega q  s_2^2 d\psi^2] \nonumber \\
&+& \g_1 \g_2   [\frac{\r^2 \o}{2}(2q(s_1^2 c_2^2
+ c_1^2 s_2^2) + \r^2 s_1^2 s_2^2) d\phi_1 d\phi_2 -\frac{1}{2}\r^2  \omega q  c_1 s_2^2 d\phi_2 d\psi - \frac{1}{2}\r^2  \omega q  c_2 s_1^2 d\phi_1 d\psi] \nonumber \\
&+&  \g_2^2 [\frac{\r^2 \o}{4}(2q(s_1^2 c_2^2 + c_1^2 s_2^2) +
\r^2 s_1^2 s_2^2) d\phi_2^2 - \frac{1}{2}\r^2  \omega q  c_2
s_1^2 d\phi_2 d\psi + \frac{1}{2}\r^2  \omega q  s_1^2 d\psi^2 ] \nonumber \\
&+& \g_1 \g_3 [\frac{\r^2 \o}{2} (2q(s_1^2 c_2^2 +
c_1^2 s_2^2) + \r^2 s_1^2 s_2^2)d\phi_1d\chi_2  -  \r^2  \omega q  c_1 s_2^2 d\chi_2d\psi] \nonumber \\
&+& \g_2 \g_3  [\frac{\r^2 \o}{2} (2q(s_1^2 c_2^2 + c_1^2 s_2^2) +
\r^2 s_1^2 s_2^2)d\phi_2d\chi_2 -\r^2 \omega q  c_2 s_1^2
d\chi_2d\psi]\}, \nonumber \eea where \bea G^{-1} &= &1 +
\g_3^2\frac{\r^2 w}{4}[2q (s_1^2 c_2^2 + c_1^2 s_2^2) + \r^2 s_1^2
s_2^2] + r^2 \sin^2\chi_1[ \g_2^2 \frac{\o}{2}  (2q c_1^2 +
\r^2 s_1^2) \nonumber \\
&+& \g_1^2 \frac{1}{2}  \o (2q c_2^2 + \r^2 s_2^2) -2\g_1 \g_2 q
\o c_1 c_2 +\g_1 \g_3 \r^2 \o f c_1 s_2^2 + \g_2 \g_3 \r^2 \o f
c_2 s_1^2].  \eea Using (\ref{3pdeformedF11asil}) we find the
deformed 4-form field $\tilde{\hat{F}}_4$ as \bea \label{mixedf}
\tilde{\hat{F}}_4 & = & \hat{F}_4 - (\frac{9q \o}{4 U})^{1/2}\r^4
s_1 s_2 d\r \wedge d\theta_1
\wedge d\theta_2 \wedge [ \g_3  d\psi ]  \nonumber \\
&+& \g_3d(G\, [\frac{\r^2 w}{4}(2q(s_1^2 c_2^2 + c_1^2 s_2^2) +
\r^2 s_1^2 s_2^2) d\phi_1 \wedge d\phi_2 \wedge d\alpha +
\frac{1}{2}\r^2 \o q c_1 s_2^2 d\phi_2 \wedge d\psi \wedge d\alpha
\nonumber \\
&-& \frac{1}{2}\r^2 \o q c_2 s_1^2 d\phi_1 \wedge d\psi \wedge
d\alpha +\frac{1}{4} \r^4 \o f s_1^2 s_2^2 d\phi_1
\wedge d\phi_2 \wedge d\psi]) \nonumber \\
&+& d\phi_1 \wedge d\psi \wedge d\chi_2 \wedge d(G\, [ \gamma_2 det\hat{g}(2,5 \mid 2,4)]) \nonumber \\
&-& d\phi_2 \wedge d\alpha \wedge d\chi_2 \wedge d(G\, [\gamma_1
det\hat{g}(1,4 \mid 1,4)
- \gamma_2 det\hat{g}(1,4 \mid 2,4) ]) \nonumber \\
&-& d\phi_1 \wedge d\phi_2 \wedge d\chi_2 \wedge d(G \, [\gamma_1
det\hat{g}(1,4 \mid 4,5)
- \gamma_2 det\hat{g}(2,4 \mid 4,5)  ]) \nonumber \\
&+& d\phi_1 \wedge d\alpha \wedge d\chi_2 \wedge d(G\, [\gamma_1
det\hat{g}(1,4 \mid 2,4)
- \gamma_2 det\hat{g}(2,4 \mid 2,4) ]) \nonumber \\
&-& d \alpha \wedge d\psi \wedge d \chi_2 \wedge d(G \, [ \gamma_1
det\hat{g}(1,2 \mid 1,4)
-\gamma_2 det\hat{g}(1,2 \mid 2,4)]) \nonumber \\
&-& d\phi_2 \wedge d\psi \wedge d\chi_2 \wedge d(G\, [ \gamma_1
det\hat{g}(1,5 \mid 1,4) ])  \, , \eea where necessary
subdeterminant are given in (\ref{subdeterminant2}). When $\g_1=\g_2=0$
the metric (\ref{mixedmetric}) and the 4-form (\ref{mixedf}) of
the mixed deformation reduce to those of a single-parameter
$\beta$ deformation (\ref{deformed3psasaki}) and (\ref{sasakif})
as expected. Similarly, when $\g_3=0$ the above metric reduces to
the metric of 2-parameter dipole deformation
(\ref{deformed3psasakidipole}) as it should. However, in this case
single $\g_1$ and $\g_2$ terms in (\ref{3dipolef}) and
(\ref{mixedf}) have opposite signs. By sending $\gamma_{1,2} \to
-\gamma_{1,2}$ in one of them, they become equal. This is not a
surprise, since we are using different orientations for the
corresponding 5-tori. In the mixed deformation our order of torus
directions is $\{\phi_1,\phi_2,\chi_2,\psi,\alpha\}$ whereas in
the dipole deformation it is
$\{\phi_1,\phi_2,\alpha,\psi,\chi_2\}$ and we have
$\epsilon^{mnp}=1$ for $m<n<p$.

\section{Conclusions and Discussions}

The main results of this paper are equations
(\ref{deformedF11asil}), (\ref{esassonuc}), (\ref{3pesassonuc})
and (\ref{3pdeformedF11asil}), along with (\ref{hatG}) and
(\ref{3phatG}). These reduce the problem of finding one or
multiparameter deformations of a  $D=11$ supergravity background
to a simple  calculation of some subdeterminants. They can be
applied to any 11-dimensional background with 5  $U(1)$
isometries, whose 4-form field has at most one leg along these.
Our method works irrespective of how the 5-torus lies in the
geometry. However, the torus coordinates should not mix with the
others. These conditions are not very restrictive, as they are met
by many frequently used M-theory solutions. Moreover, our results
can be adopted easily to backgrounds with only four or three
$U(1)$ directions, as we showed in (\ref{4U1}), (\ref{3U1}) and
(\ref{4U12}). We also explained in section 2.3, through a specific
example, how our method is modified, when the  condition on the
4-form field is violated. Naturally, our results can be applied to
many other interesting backgrounds. For instance, the dipole or
multiparameter deformations of the solutions considered in
\cite{ahn, jerome} can be obtained easily. We hope that our
formulas will be useful in the construction of such new examples,
especially for multiparameter deformations.

Although our method works for any M-theory background with 5
$U(1)$ isometries\ we demonstrated our results with backgrounds of
the form $AdS_4 \times M_7$ or $AdS_7 \times M_4$, as they are of
obvious interest for the AdS/CFT correspondence. In the first
case, the dual field theory can be regarded as a three dimensional
field theory arising on the world-volume of coincident M2 branes,
or more appropriately, as the IR limit of the field theory on the
world-volume of coincident D2 branes, from the IIA perspective.
The Sasaki-Einstein manifolds we consider  in this paper have two
Killing spinors, and hence the dual field theory is an
$\mathcal{N}$=2 supersymmetric field theory. They have large
isometry groups ($SU(2) \times SU(2) \times U(1)^2$ for the one
with the
 base $S^2 \times S^2$ and $SU(3) \times U(1)^2$
when the base is $CP^2$), which correspond to the global
symmetries of the dual field theory. In each case, the Killing
spinors transform as a \textbf{2} of one $U(1)$
factor\footnote{Here we identify $U(1) \sim SO(2)$.} of the
isometry group and this corresponds, on the field theory side, to
the $U(1)$ R-symmetry, which acts on the supercharges. Therefore,
we expect that the dual field theory remains $\mathcal{N}$=2
supersymmetric, as long as this particular $U(1)$ (whose
corresponding Killing vector is the Reeb vector) is not involved
in the deformation process, which was verified explicitly for our
one parameter $\beta$ deformation case in \cite{jerome}. Hence,
we expect our examples in sections 2.1 and 3.1 to preserve $\mathcal{N}$=2 supersymmetry
except the 3-parameter $\beta$
deformation case discussed in section 3.1.1. This last one
should break supersymmetry
completely, as any symmetry that leaves one of the Killing spinors
invariant should also leave the other invariant (for a general
argument, see \cite{jerome}). For the $AdS_7 \times S^4$ case the
dual theory is a six dimensional $\mathcal{N}$=(2,0)
supersymmetric CFT \cite{mal1}. We expect supersymmetry to be
preserved in our noncommutative deformation since the $SO(5)$
R-symmetry remains intact, whereas it should be broken in dipole
deformations.

An important problem  here is to identify the marginal operators
(for the $\beta$ deformations) on the field theory side, that
corresponds to our deformations. Let us recall 
the $AdS_5 \times S^5$ example, whose dual field theory is
$d=4$, $\mathcal{N}$=4 supersymmetric Yang-Mills theory.
Here, choosing all the deformation directions from the $AdS_5$ part corresponds to
a noncommutative deformation on the field theory side, whereas
choosing one direction from the $AdS_5$ and one from the $S^5$ results
in a dipole deformation.  Making analogy with the noncommutative case, Lunin and
Maldacena showed that, choosing both $U(1)$'s from the $S^5$
part should give the duals of $\beta$ deformations of the
$\mathcal{N}$=4 theory.  The effect in the Lagrangian is to modify
the product of fields charged under the global $U(1) \times U(1)$,
just like the noncommutative deformations modify the ordinary
product to a star product. This introduces phases in the
Lagrangian that depend on the deformation parameter (and all three
parameters for the 3-parameter case, see \cite{frolov}). This
argument does not carry over directly to the 3 dimensional CFTs
\cite{jerome}, that arise as duals of the $AdS_4 \times M_7$
backgrounds, although some steps have been taken in this direction
\cite{berman2}. An alternative way to find the marginal operators
for the dual field theory is to notice that the deformation, for
small values of the deformation parameter, corresponds to turning
on a massless mode in the Kaluza-Klein spectrum of the undeformed
background. Then the dual operator should be of dimension $d$,
where $d=4$ for the $AdS_5$ case and $d=3$ for the $AdS_4$. Combining
this with the fact that it has to belong to a short-multiplet and
break the global symmetry group to $U(1)^3$ (which is because the
isometry group is broken to $U(1)^3$ on the gravity side), one can
identify the marginal operator in question. In this way, Lunin and
Maldacena determined the marginal operator, which should
correspond to the $\beta$ deformation of $AdS_4 \times S^7$
\cite{mal2}.  Similarly, \cite{jerome} proposed the operators
corresponding to the deformations of the Sasaki-Einstein manifolds
$M(3,2)$ and $Q(1,1,1)$. These two Sasaki-Einstein manifolds are
special in that they have proposals for their field theory duals
\cite{fabri}. We expect that similar arguments can be made for the
Sasaki-Einstein manifolds we consider here, once their field
theory duals (before deformation) is understood better. Recently,
there have been important developments in this direction
\cite{ek1}. The duals of the mixed deformations would be especially interesting to study, 
since they have not been analyzed before elsewhere.

There are two straightforward generalizations of our results. One
is to consider geometries with more than five decoupled $U(1)$
directions or allow some coordinates to mix with these five
$U(1)$'s. Another is to construct deformations with more than 3
parameters. Actually, as we saw when the number of $U(1)$
directions is $n$, there can be $n!/6(n-3)!$ parameters. Sometimes
one can have even supersymmetric deformations with more than 3
parameters. For example, in the mixed deformation case above, it
is possible to have a 4-parameter deformation without using the
R-symmetry direction.

It would  also be interesting to study giant gravitons
\cite{susskind} on our new backgrounds. Giants on 10-dimensional
$\beta$ deformed solutions were analyzed in \cite{pir, naq, butti,
pirrone2}. It is desirable to extend these to $D=11$ and to other
types of deformations, which we aim to study in the near future.

\

\noindent {\bf Acknowledgements}

NSD is partially supported by Turkish Academy of Sciences via The
Young Scientists Award Program (T\"UBA- GEB\.IP). He also wishes
to thank the Abdus Salam ICTP for hospitality where some part of
this paper was written.

\

\appendix

\section{Proof I}
\label{A}

Here we will prove the equivalence of (\ref{deformedF11}) and
(\ref{deformedF11asil}). First let us compare the second terms by
using  the following fact: If we make a dimensional reduction from
$(D+1)$ to $D$ dimensions by using the ansatz
$$ds_{D+1}^2 = e^{2\alpha \phi}ds_D^2 + e^{2\beta \phi}(dz+A)^2 \, , $$
then for an $n-$form $X_n$ we have the following relations
\cite{fibre}: \bea \label{hodge} \star_{(D+1)}X_n &=&
e^{[(D-2n)\alpha +
\beta]\phi}(-1)^n \star_{D} X_n \wedge (dz +A),  \\
\label{star2} \star_{(D+1)}[X_n \wedge (dz +A)] &=&
e^{[(D-2n)\alpha - \beta]\phi} \star_{D} X_n,  \eea where
$\star_{(D+1)}$ and $\star_{D}$ denote the Hodge duals taken in
$D+1$ and $D$ dimensions, respectively. In our case $D = 10, n=4,
\alpha = -1/3$ and $\beta = 2/3$. Therefore, we have
$$\star_{11} F_4 = \star_{10} F_4 \wedge (dz+A).$$ \noindent As a
result we see that $$i_{\partial/\partial z}\star_{11}F_4 =
\star_{10} F_4.$$ \noindent On the other hand, it easily follows
from (\ref{star2}) that $i_{\partial/\partial z}\star_{11}[F_3
\wedge (dz +A)] = 0$. Thus,
$$i_{\partial/\partial z}\star_{11}\hat{F}_4 = i_{\partial/\partial z}\star_{11}[F_4
+ F_3 \wedge (dz +A)] = \star_{10} F_4.$$ This shows the equality
of the second terms in (\ref{deformedF11}) and
(\ref{deformedF11asil}).

Now we compare the third terms of (\ref{deformedF11}) and
(\ref{deformedF11asil}). There are 10 terms to be
 compared. The equality of the coefficients of the 6 of these terms which are of the form $dx_i
 \wedge dx_j \wedge dz$ is obvious. From (\ref{deformedF11}) we
 see that the coefficients are $d\tilde{B}_{ij}$ and reading these
 terms from (\ref{compactB}) we can directly observe the equality. Now let us look at the coefficients of
 the $dx_1 \wedge dx_2 \wedge dx_3$ terms. From (\ref{compactB}), (\ref{deformedF11}) and
(\ref{equality}) we read the coefficient as \be \label{123} \gamma
G[det\hat{g}(3,4 \mid 3,4) A_3 - det\hat{g}(2,4 \mid 3,4) A_2 +
det\hat{g}(1,4 \mid 3,4) A_1]. \ee Using the fact that $A_i =
\hat{g}_{iz}/\hat{g}_{zz}$ (this can be seen directly from
(\ref{metric11a})) one can make the following observation: If we
subtracted from (\ref{123}) the term $\gamma G[det\hat{g}(3,4 \mid
z,4)]$ then the result would be  $\gamma G$ times the determinant
of a new matrix, say $K$, obtained from $\hat{g}(4 \mid 3)$ by
replacing its third row by the row $\{\hat{g}_{1z}/\hat{g}_{zz},
\hat{g}_{2z}/\hat{g}_{zz}, \hat{g}_{3z}/\hat{g}_{zz},1 \}$. It is
easy to see that $det K = 0$ as its third row is a
$1/\hat{g}_{zz}$ multiple of its fourth row. Therefore we conclude
that (\ref{123}) is equal to $\gamma G(det\hat{g}(3,4 \mid z,4))$,
which is exactly the term that is given by the third term in the
formula (\ref{deformedF11asil}).

One can also show that the coefficients of the terms $dx_1 \wedge
dx_2 \wedge dx_4$, $dx_1 \wedge dx_3 \wedge dx_4$ and $dx_2 \wedge
dx_3 \wedge dx_4$ which are read from (\ref{deformedF11}) and
(\ref{deformedF11asil}) are equal by using arguments similar to
the above.

\section{Proof II}
\label{B}

Here we will prove equation (\ref{equality}). To do this we define
a new matrix $\hat{g}'$ as the $5 \times 5$ matrix with entries
$(\hat{g}')_{ij} = e^{-2/3 \phi}(g)_{ij} = (\hat{g}_{zz})^{-1/2}
(g)_{ij}$, \ \ $(\hat{g}')_{i5} = (\hat{g}')_{5i} = 0, \ \ i,j =
1,2,3,4$ and $(\hat{g}')_{55} = e^{4/3 \phi} = \hat{g}_{zz}$. This
new matrix $\hat{g}'$ and the original $5 \times 5$ matrix
$\hat{g}$ with entries $(\hat{g})_{ab} = \hat{g}_{ab},  \ a,b =
1,2,3,4,z$ are related under the following transformation \be
\label{similarity} \hat{g} = S^T \ \hat{g}' \ S, \ee where \be
\label{S} S = \left(\begin{array}{ccccc} 1 & 0 &
0 & 0 & 0 \\
0 & 1 & 0 & 0 & 0 \\
0 & 0 & 1 & 0 & 0 \\
0 & 0 & 0 & 1 & 0 \\
A_1 & A_2 & A_3 & A_4 & 1
\end{array}\right), \ee
where $A_i$ are the functions appearing in $A = A_i dx^i,
i=1,2,3,4$. This can be seen directly from (\ref{metric11a}).
Therefore, $det(\hat{g}) = det(S)^2 det(\hat{g}') =
det(\hat{g}')$. Similarly one can show that \bea
\label{similarity2} \hat{g}(i \mid j) &=& S^T(j \mid j) \
\hat{g}'(i
\mid j) \ S(i \mid i), \\
 \label{similarity3} \hat{g}(i,j
\mid k,l) &=& S^T(k,l \mid k,l) \ \hat{g}'(i,j \mid k,l) \ S(i,j
\mid i,j), \eea which implies \be \label{det1} det \hat{g}(i \mid
j) = det \hat{g}'(i \mid j), \ \ \ \ det \hat{g}(i,j \mid k,l) =
det \hat{g}'(i,j \mid k,l). \ee On the other hand, from the the
way we have defined the matrix $\hat{g}'$ (\ref{similarity}) we
see that \bea \label{equality2}  det \hat{g}'(i \mid j) &=&
\hat{g}_{zz} (\hat{g}_{zz})^{-3/2} detg(i \mid j) =
(\hat{g}_{zz})^{-1/2}detg(i \mid j) \nonumber \, , \\
det \hat{g}'(i,j \mid k,l) &=& \hat{g}_{zz} (\hat{g}_{zz})^{-1}
detg(i,j \mid k,l) = detg(i,j \mid k,l), \eea where we have used
$e^{-2/3 \phi} = (\hat{g}_{zz})^{-1/2}$ and $\hat{g}'(i \mid j)$
and $\hat{g}'(i,j \mid k,l)$ are $3 \times 3$ and $2 \times 2$
matrices, respectively. (\ref{det1}) and (\ref{equality2})
together prove (\ref{equality}).

\section{Subdeterminants}
The relevant non-zero subdeterminants of the matrix (\ref{bgs2s2}) are:
\bea
det\hat{g}(4 \mid 4)& = & \frac{\omega q
\r^4
s_{1}^2s_{2}^2}{4} \nonumber\\
det\hat{g}(3 \mid 3)& = &\frac{1}{4}\r^2 r^2 \omega \sin^2\chi_1 [2 q (s_1^2 c_2^2 + s_2^2 c_1^2) + \r^2 s_1^2 s_2^2] \nonumber \\
det\hat{g}(2 \mid 2)& = & \frac{1}{2}\r^2 r^2 \omega q \sin^2\chi_1 s_1^2\nonumber\\
det\hat{g}(1 \mid 1)& = & \frac{1}{2}\r^2 r^2 \omega q \sin^2\chi_1 s_2^2  \nonumber\\
det\hat{g}(1 \mid 3)& = &  \frac{1}{2}\r^2 r^2 \omega q \sin^2\chi_1 c_1 s_2^2 \nonumber\\
det\hat{g}(2 \mid 3)& = & -\frac{1}{2}\r^2 r^2 \omega q \sin^2\chi_1 c_2 s_1^2 \nonumber\\
det\hat{g}(3,4 \mid 3,4) &=&
\frac{\r^2 \o}{4}[2q(s_1^2 c_2^2 +
c_1^2 s_2^2) + \r^2 s_1^2 s_2^2] \nonumber\\
det\hat{g}(2,4 \mid 2,4) &=&\frac{1}{2}\r^2 \o q s_1^2 \nonumber\\
det\hat{g}(1,4 \mid 1,4) &=& \frac{1}{2}\r^2 \o q s_2^2  \nonumber\\
det\hat{g}(1,4 \mid 3,4) &=& \frac{1}{2}\r^2 \o q c_1 s_2^2 \nonumber\\
det\hat{g}(2,4 \mid 3,4) &=& -\frac{1}{2}\r^2 \o q c_2 s_1^2 \nonumber\\
det\hat{g}(4,5 \mid 3,4) &=& \frac{1}{4} \r^4 \o f s_1^2 s_2^2.
\label{subdeterminant1}
\eea

The relevant non-zero subdeterminants of the matrix (\ref{bgs2s2dipole}) are:

\bea
det\hat{g}(4 \mid 4)& =
& \frac{1}{4} r^2 \sin^2\chi_1 \r^2 \o[2q(s_1^2 c_2^2 +
c_1^2 s_2^2) + \r^2 s_1^2 s_2^2] \nonumber\\
det\hat{g}(3 \mid 3)& = &\frac{1}{4} r^2 \sin^2\chi_1 \r^4   (q + \o f^2)  s_1^2 s_2^2 \nonumber \\
det\hat{g}(2 \mid 2)& = & \frac{1}{2}r^2 \sin^2\chi_1  \r^2  \omega q  s_1^2\nonumber \\
det\hat{g}(1 \mid 1)& = & \frac{1}{2}r^2 \sin^2\chi_1  \r^2  \omega q  s_2^2 \nonumber\\
det\hat{g}(1 \mid 4)& = &  -\frac{1}{2}r^2 \sin^2\chi_1  \r^2  \omega q  c_1 s_2^2 \nonumber\\
det\hat{g}(2 \mid 4)& = & \frac{1}{2}r^2 \sin^2\chi_1 \r^2  \omega q  c_2 s_1^2 \nonumber\\
det\hat{g}(3 \mid 4)& = & \frac{1}{4} r^2 \sin^2\chi_1 \r^4 \o f
s_1^2 s_2^2 \nonumber\\
det\hat{g}(3,4 \mid 3,4) &=& \frac{1}{4} r^2
\sin^2\chi_1 \r^2 [2(q + \o f^2)(s_1^2 c_2^2 +
c_1^2 s_2^2) + \r^2 s_1^2 s_2^2] \nonumber\\
det\hat{g}(2,4 \mid 2,4) &=&\frac{1}{2} r^2 \sin^2\chi_1  \o (2q c_1^2 + \r^2 s_1^2) \nonumber\\
det\hat{g}(1,4 \mid 1,4) &=& \frac{1}{2}r^2 \sin^2\chi_1  \o (2q c_2^2 + \r^2 s_2^2)  \nonumber\\
det\hat{g}(1,4 \mid 3,4) &=& \frac{1}{2}r^2 \sin^2\chi_1  \r^2 \o f c_1 s_2^2 \nonumber\\
det\hat{g}(2,4 \mid 3,4) &=& -\frac{1}{2}r^2 \sin^2\chi_1 \r^2 \o f c_2 s_1^2 \nonumber\\
det\hat{g}(1,4 \mid 2,4) &=& r^2 \sin^2\chi_1 q \o c_1 c_2 \nonumber\\
det\hat{g}(1,3 \mid 3,4) &=& \frac{1}{2}r^2 \sin^2\chi_1 \r^2 (q + \o f^2) c_1 s_2^2 \nonumber\\
det\hat{g}(2,3 \mid 3,4) &=& -\frac{1}{2}r^2 \sin^2\chi_1 \r^2  (q + \o f^2) c_2 s_1^2 \nonumber\\
det\hat{g}(1,2 \mid 2,4) &=& r^2 \sin^2\chi_1 q \o c_1  \nonumber\\
det\hat{g}(1,2 \mid 1,4) &=& r^2 \sin^2\chi_1 q \o  c_2 \nonumber\\
det\hat{g}(2,3 \mid 2,4) &=& \frac{1}{2}r^2 \sin^2\chi_1 \r^2 \o f s_1^2 \nonumber\\
\label{subdeterminant} det\hat{g}(1,3 \mid 1,4) &=& \frac{1}{2}r^2
\sin^2\chi_1 \r^2 \o f s_2^2.
\eea

The relevant non-zero subdeterminants of the matrix (\ref{bgs2s2mixed}) are:
\bea
det\hat{g}(4 \mid 4)& = & \frac{\r^2 \o}{4}
r^2\sin^2\chi_1[2q(s_1^2 c_2^2 +
c_1^2 s_2^2) + \r^2 s_1^2 s_2^2] \nonumber\\
det\hat{g}(3 \mid 3)& = &\frac{1}{4}\r^4  q \omega   s_1^2 s_2^2 \nonumber \\
det\hat{g}(2 \mid 2)& = & \frac{1}{2} r^2 \sin^2\chi_1 \r^2  \omega q  s_1^2\nonumber \\
det\hat{g}(1 \mid 1)& = & \frac{1}{2}r^2 \sin^2 \chi_1 \r^2  \omega q  s_2^2 \nonumber\\
det\hat{g}(1 \mid 4)& = &  -\frac{1}{2} r^2 \sin^2 \chi_1  \r^2  \omega q  c_1 s_2^2 \nonumber\\
det\hat{g}(2 \mid 4)& = & \frac{1}{2} r^2 \sin^2 \chi_1  \r^2
\omega q  c_2 s_1^2  \nonumber\\
det\hat{g}(3,4 \mid 3,4) &=&
\frac{\r^2 w}{4}[2q (s_1^2 c_2^2 +
c_1^2 s_2^2) + \r^2 s_1^2 s_2^2] \nonumber\\
det\hat{g}(2,4 \mid 2,4) &=&\frac{1}{2} r^2 \sin^2\chi_1 \o (2q c_1^2 + \r^2 s_1^2) \nonumber\\
det\hat{g}(1,4 \mid 1,4) &=& \frac{1}{2}r^2 \sin^2\chi_1 \o (2q c_2^2 + \r^2 s_2^2)  \nonumber\\
det\hat{g}(1,4 \mid 2,4) &=& r^2 \sin^2\chi_1 q \o c_1 c_2 \nonumber\\
det\hat{g}(1,3 \mid 3,4) &=& \frac{1}{2}\r^2 q \o  c_1 s_2^2 \nonumber\\
det\hat{g}(2,3 \mid 3,4) &=& -\frac{1}{2}\r^2 q \o  c_2 s_1^2 \nonumber\\
det\hat{g}(1,2 \mid 2,4) &=& r^2 \sin^2\chi_1 q \o c_1  \nonumber\\
det\hat{g}(1,2 \mid 1,4) &=& r^2 \sin^2\chi_1 q \o  c_2 \nonumber\\
det\hat{g}(4,5 \mid 1,4) &=& -\frac{1}{2} r^2 \sin^2\chi_1 \r^2 \o
f c_1 s_2^2 \nonumber \\
det\hat{g}(4,5 \mid 2,4) &=& \frac{1}{2} r^2 \sin^2\chi_1 \r^2 \o
f c_2 s_1^2 \nonumber \\
det\hat{g}(2,5 \mid 2,4) &=& \frac{1}{2} r^2 \sin^2\chi_1 \r^2 \o
f  s_1^2 \nonumber \\
det\hat{g}(1,5 \mid 1,4) &=& \frac{1}{2} r^2 \sin^2\chi_1 \r^2 \o
f  s_2^2 \nonumber \\
\label{subdeterminant2} det\hat{g}(3,5 \mid 3,4) &=& \frac{1}{4}
\r^4 \o f s_1^2 s_2^2 \, .
\eea

\end{document}